\documentclass[reprint,amsmath,amssymb,aps,prb]{revtex4-2}
\usepackage{multirow}
\usepackage{graphicx}
\usepackage{dcolumn}
\usepackage{comment}
\usepackage{float}
\usepackage{bm}
\usepackage{booktabs}
\usepackage[colorlinks=true, allcolors=blue]{hyperref}

\begin{document}

\preprint{CSG-2025-1}

\title{Machine-learning semi-local exchange-correlation functionals for Kohn-Sham density functional theory of 
the Hubbard model}

\author{Eoghan Cronin$^1$}
\email{eoghanc@tcd.ie}

\author{Rajarshi Tiwari$^{1,2}$}
\email{rajarshi.tiwari@ichec.ie}

\author{Stefano Sanvito$^1$}
\email{sanvitos@tcd.ie}

\affiliation{
 ${}^1$ School of Physics, AMBER and CRANN Institute, Trinity College, Dublin 2, Ireland
\\
${}^2$ Irish Centre for High-End Computing (ICHEC),
    University of Galway, Ireland.
}

\date{\today}

\begin{abstract}

The Hubbard model provides a test bed to investigate the complex behaviour arising from 
electron-electron interaction in strongly-correlated systems and naturally emerges as the 
foundation model for lattice density functional theory (DFT). Similarly to conventional DFT, 
lattice DFT computes the ground-state energy of a given Hubbard model, by minimising a universal 
energy functional of the on-site occupations. Here we use machine learning to construct a class 
of scalable `semi-local' exchange-correlation functionals with an arbitrary degree of non-locality 
for the one-dimensional spinfull Hubbard model. Then, by functional derivative we construct an 
associated Kohn-Sham potential, that is used to solve the associated Kohn-Sham equations. After 
having investigated how the accuracy of the semi-local approximation depends on the degree 
of non-locality, we use our Kohn-Sham scheme to compute the polarizability of linear chains,
either homogeneous or disordered, approaching the thermodynamic limit.
\end{abstract}

\maketitle

\section{\label{sec:level1}Introduction}
Over the last few decades, density functional theory (DFT) has become the computational 
method of choice across several different fields, going from quantum chemistry to material 
science. Today, DFT serves as the primary tool for calculating the electronic structure of 
both solids and molecules, thanks to its foundation in the rigorously established 
Hohenberg-Kohn theorems~\cite{HK1964}, to the practical Kohn-Sham functional-minimization 
scheme~\cite{KS1965}, and to the numerous available efficient numerical 
implementations~\cite{Lejaeghere2016}. The Hohenberg-Kohn theorems
establish the existence of a universal energy functional, $F[n]$, of the electron charge 
density, $n(\mathbf{r})$, functional that presents its minimum at the ground-state density. 
Thus, in principle, one can solve any ground-state electronic quantum-mechanics problem, 
by simply scanning through all the legitimate charge densities, in the search of the functional 
minimum. This conceptually simple strategy, known as orbital-free DFT~\cite{OFDFT}, is 
however practically impossible. 

Two main obstacles prevent the direct minimization of $F[n]$. Firstly, the exact functional 
is unknown, so that an exact solution of the problem is not available. This is partially 
mitigated by the possibility to break the functional into three contributions, (i) the kinetic 
energy of a non-interacting system having the same density, $T_0[n]$, (ii) the classical 
Coulomb energy (Hartree energy) and (iii) the so-called exchange-correlation energy, $E_\mathrm{xc}[n]$. 
This last term describes the non-classical electron-electron interaction and the correction to the 
non-interacting kinetic energy arising from electron correlation. Importantly, there are many 
approximations to $E_\mathrm{xc}[n]$ that allows one to perform accurate calculations. Hence, 
the DFT problem, at least in an approximated form, is solvable. Yet, orbital-free DFT remains 
challenging, since the exact density-functional form for the non-interacting kinetic energy is 
unknown and there are no accurate and universal approximations for it working across different 
external potentials.

As a consequence, practically any numerical implementation of DFT follows the Kohn-Sham (KS) 
procedure. This consists in mapping the interacting many-body problem onto a fictitious single-particle 
one where, by construction, the two problems share the same ground-state energy and density. The 
potential associated to the single-particle problem is constructed from $F[n]$ by functional derivative 
and it is itself density dependent. As such, the non-interacting kinetic energy is computed
exactly, but the solution of the KS equations must be iterative. Although, in principle, the 
exact functional depends on the electron density everywhere in space, namely it is non-local, 
the most common approximations to $E_\mathrm{xc}[n]$ are local or semilocal. For instance, the 
local density approximation (LDA) assigns the exchange-correlation energy at a given point to 
that of the homogeneous electron gas computed for the density at that point \cite{KS1965}. More 
sophisticated and accurate approximations can then be constructed by expanding the functional 
dependence of the exchange-correlation energy to the density gradient and higher derivatives, 
following a systematic Jacob's ladder \cite{PS2001}. These work well across a broad range of 
electronic-structure types, but they usually fail for systems presenting strongly correlated 
electrons \cite{MS2009}. The systematic construction of non-local functionals that satisfy the 
known conditions of exact DFT is highly non-trivial and their deployment in calculations for 
real systems is extremely numerically demanding. 

Some of these challenges motivated the construction of DFT for lattice models, namely of lattice 
density functional theory (LDFT)~\cite{GS1986,SGN1995}. To fix the idea, let us consider the 
one-dimensional (1D) single-orbital Hubbard model~\cite{Hubbard1964}. LDFT then translates the 
Hohenberg-Kohn theorems by replacing the electron density with the site occupation. One can then 
define a universal functional that depends on the entire occupation manifold, namely on the 
average occupation of each site comprising the model. Although one should be careful to transfer 
DFT concepts to LDFT~\cite{Capelle2013,Franca2018,Penz2021,Sobrino2023}, the study of lattice 
models has a number of advantages. In particular, in LDFT the charge density has a simple single
representation, some exact limiting solutions are known~\cite{HubbardBook}, and, for relatively 
small systems, exact numerical results can be obtained at a moderate computational cost. As such, 
LDFT can be seen as a simple playground to investigate some fundamental questions of DFT. For 
instance, by using the Bethe ansatz solution of the Hubbard model one can construct LDA 
functionals~\cite{Lima2002,Lima2003} and then investigate the response of strongly correlated systems 
to electric~\cite{Akande2010} and magnetic fields~\cite{Akande2012}, their susceptibility~\cite{Schenk2008}, 
the formation of the Mott gap~\cite{Franca2012} and quantum transport~\cite{Vettchinkina2013}.

An interesting aspect of any functional theory is given by the possibility of combining it 
with machine-learning (ML) techniques. In fact, within conventional DFT, ML appears ideal to 
construct numerical mappings between the electron density and the 
energy~\cite{Snyder2012,Li2016a,Li2016b,Mills2017,Schmidt2019,Ryczko2019,Kalita2021,Nunez2021}, 
between the external potential and the electron density~\cite{Brockherde2017,Focassio2023,Focassio2024}, 
between the electron density and the ground-state wave function~\cite{Moreno2020} or between the 
electron density and an observable~\cite{Moreno2021}. Moving to lattice models, we have shown that 
a numerically exact functional can be constructed for the spin-full Hubbard model in one 
dimension~\cite{Nelson2019}. This depends on the site occupations of the entire system, namely it 
is a fully non-local functional, and it satisfies both Hohenberg-Kohn theorems. Unfortunately, since 
the energy depends on the entire site occupation manifold, a functional trained on a $N$-site 
system cannot be used to predict the energy of systems of other size. We have then avoided such 
drawbacks by constructing semi-local ML functionals, this time for the spinless Hubbard 
model~\cite{Nelson2021}. In both of these two examples, the energy minimization is performed in 
an orbital-free-DFT fashion. 

In this work, we take an alternative approach, namely we demonstrate that a Kohn-Sham potential 
for the spin-full 1D Hubbard model can be obtained from the numerical ML energy functional by 
functional derivative. This is constructed semi-locally so that it can be trained on relatively small 
systems and then used for much larger ones. With this, we are able to investigate relevant quantities as 
they approach the thermodynamic limit, namely for a large number of sites. In particular, we 
concentrate here on the charge density response of the one-dimensional Hubbard model in the 
metallic state and in the presence of disorder, for which we evaluate the polarizability.

The paper is structured as follows. In Section~\ref{method} we present our methodology,
by setting up the problem, discussing how to generate the target data set, and by explaining 
how we train the ML functional. Our presentation of the results begins in Section~\ref{3a} 
by demonstrating the performance of our KS equations and the accuracy of the constructed 
KS potential. This is followed by an analysis of the non-locality of the functional in 
Section~\ref{a_dependence}. We then proceed to look at some applications of the trained functional.
In Sec.~\ref{extrapolation_section} we apply it to solve systems with a filling factor outside 
the range used for the training. Then, we analyze the (deviation from) piecewise-linearity of the
energy at fractional electron filling and the associated derivative discontinuity [see Section 
\ref{piecewise_linearity_section}]. Finally, in Section~\ref{polarizability_section} we investigate 
the response of the charge density to an electric field and the ability of the model to solve 
systems with thousands of lattice sites. Conclusions are drawn in Section~\ref{conclusions}.

\section{\label{method}Method}

\subsection{\label{setup}Kohn-Sham LDFT setup for Hubbard model}

Our analysis is based on the one-dimensional, single-orbital, spin-full Hubbard model~\cite{Hubbard1964}, 
described by the following Hamiltonian,
\begin{equation}\label{HubbardU}
\hat{H}_U = \hat{T} + \hat{U} + \sum_{i\sigma} v_i \hat{n}_{i\sigma}\:.
\end{equation}
This comprises the kinetic energy, 
\begin{equation}\label{KE}
\hat{T} = -t \sum_{i\sigma}\left(\hat{c}_{i \sigma}^{\dagger} \hat{c}_{i+1,\sigma}+ \hat{c}_{i+1,\sigma}^{\dagger} \hat{c}_{i \sigma}\right)\:,
\end{equation}
the electron-electron repulsion,
\begin{equation}
\hat{U} = U \sum_{i} \hat{n}_{i\uparrow} \hat{n}_{i\downarrow}\:,
\end{equation}
and the external potential, defined through the set of on-site energies, $\{v_i\}$. Here 
$\hat{c}_{i \sigma}^{\dagger}$ ($\hat{c}_{i \sigma}$) is the fermionic creation (annihilation) 
operator for an electron at site $i$ with spin $\sigma=\uparrow, \downarrow$, the
$\hat{n}_{i\sigma} = \hat{c}_{i\sigma}^{\dagger}\hat{c}_{i\sigma}$ are the occupation operators,
while $t$ and $U>0$ are the electronic hopping integral and the Coulomb repulsion energy,
respectively.

The LDFT version of the Hohenberg-Kohn theorems~\cite{GS1986,SGN1995} establishes that there exists 
a universal functional, $F_U\left[\left\{n_{i \sigma}\right\}\right]$, of the spin-resolved site 
occupations, $n_{i \sigma}=\langle\hat{n}_{i\sigma}\rangle$, such that the total energy of the system,
$E_{U}\left[\left\{n_{i \sigma}\right\}\right]$, can be written as 
\begin{equation}\label{E_U}
E_{U}\left[\left\{n_{i \sigma}\right\}\right]=F_U\left[\left\{n_{i \sigma}\right\}\right]+\sum_{i \sigma} n_{i \sigma} v_{i}\:,
\end{equation}
where by definition,
\begin{equation}\label{F_U}
F_{U}\left[\left\{n_{i \sigma}\right\}\right]=\langle\Psi\left[\left\{n_{i \sigma}\right\}\right]|\hat{T}+\hat{U}| \Psi\left[\left\{n_{i \sigma}\right\}\right]\rangle\:,
\end{equation}
with $| \Psi\left[\left\{n_{i \sigma}\right\}\right]\rangle$ being the many-body wave function. 
Note that the functional depends on the Coulomb repulsion energy $U$ (there is a different universal
functional for every different $U$), and the $U=0$ limit recovers the non-interacting tight-binding model.
Thus, in analogy with conventional DFT, one defines the exchange-correlation energy $E_\mathrm{xc}[\left\{n_{i \sigma}\right\}]$, as
\begin{equation}
    E_\mathrm{xc}[\left\{n_{i \sigma}\right\}] =F_U\left[\left\{n_{i \sigma}\right\}\right] - F_{U=0}\left[\left\{n_{i \sigma}\right\}\right]\:,
    \label{ehxc1}
\end{equation}
where, at variance with DFT, we have not singled out the classical Hartree contribution. With such 
definition at hand we can readily write down the associated KS equations as follows
\begin{equation}
    \big(\hat{T}+{v}_{\text{eff}, i\sigma}\big) \phi_{m}=\epsilon_{m} \phi_{m},
\end{equation}
where ${v}_{\text{eff}, i\sigma} = v_{\mathrm{xc}, i\sigma}+ v_i$ is the KS potential, while $\epsilon_m$ 
and $\phi_m$ are the KS energies and orbitals, respectively. Formally, the exchange-correlation potential 
is obtained as functional derivative of the corresponding energy, 
$v_{\mathrm{xc}, i\sigma}=\frac{\partial E_\mathrm{xc}}{\partial n_{i \sigma}}$, but in practice this 
is just a standard derivative, since $E_\mathrm{xc}$ is a function of the $2L$-dimensional vector of 
site occupations, $\{n_{i\sigma}\}$, where $L$ is the total number of sites.

Throughout this work, we will consider the case of one-third filling, $N_\mathrm{e}/L = 2/3$ 
($N_\mathrm{e}$ is the total number of electrons), which is representative of the metallic phase. 
Thus, we stay away from half-filling, corresponding to the Mott insulating phase, as this presents 
a much longer correlation length and it is usually much harder to converge. Furthermore, we look 
at finite-size chains with open boundary conditions and a net spin-zero state, namely at the situation 
$N_{\uparrow} = N_{\downarrow}=N_\mathrm{e}/2$. In this case, spin symmetry allows us to drop the spin 
index and we will express all quantities in terms of the total occupation $n_i = n_{i\uparrow} + n_{i\downarrow}$.

\subsection{\label{data_generation_section}Generation of the training and test sets}
The construction of the ML exchange-correlation energy consists in establishing a mapping between the 
ground-state site occupations, $\left\{n_{i \sigma}\right\}$, and $E_\mathrm{xc}$. As such, for a given 
set of on-site energies $\left\{v_{i}\right\}$ one needs to compute the exact ground-state energy of 
the system. The value of $E_\mathrm{xc}$ is then obtained by subtracting the contribution from the 
external potential [see Eq.~(\ref{HubbardU})] and the kinetic energy of the non-interacting system
[see Eq.~(\ref{ehxc1})]. We use the density-matrix renormalization group (DMRG) \cite{White1992} algorithm 
via TenPy \cite{tenpy} to solve the full many-body problem. Throughout this work, we fix the hopping parameter 
to $t=1$, which sets the energy scale of the problem, and consider the case $U=4$. The training set is 
constructed from data corresponding to lattice of different sizes, $L = 18, 21, 24$, and for each one 
of them we produce 17,000 data points.

In order to generate a dense and diverse training dataset, we use the following strategy. A data 
point here corresponds to a different choice of external potential, generated randomly within an 
interval $v_i^{(k)} \in [0,\lambda^{(k)}]$. In particular, for each lattice size, $n=15,000$ configurations 
are obtained from $\lambda^{(k)} = \sqrt{9k/(n-1)}$. The first data point $\lambda^{(k=0)} = 0$ gives us
the homogeneous case, while the final point gives $\lambda^{(k=14999)} = 3$. The square root dependence 
on $k$ is selected as it gives rise to a good distribution of densities, as it pushes more of the data 
points further away from the homogeneous case. Then, we obtain a further $n=2,000$ configurations from 
$\lambda^{(k)} = 3+3k/(n-1)$, a choice that guarantees enough diversity in the training set. Finally, 
we also include in the training dataset potentials obtained by applying constant electric fields of 
different strength [see Section \ref{polarizability_section}] over the homogeneous lattice $v_i=0$. 
The DMRG data are here deemed as `exact', although their accuracy is limited by the DMRG calculation 
itself. Poorly converged DMRG data are not included in the training set. The threshold of accuracy was 
set by requiring that the relative change in the energy is within $10^{-5}$ after the final step. 

\begin{figure}[t]
    \centering
    \includegraphics[scale = 0.25]{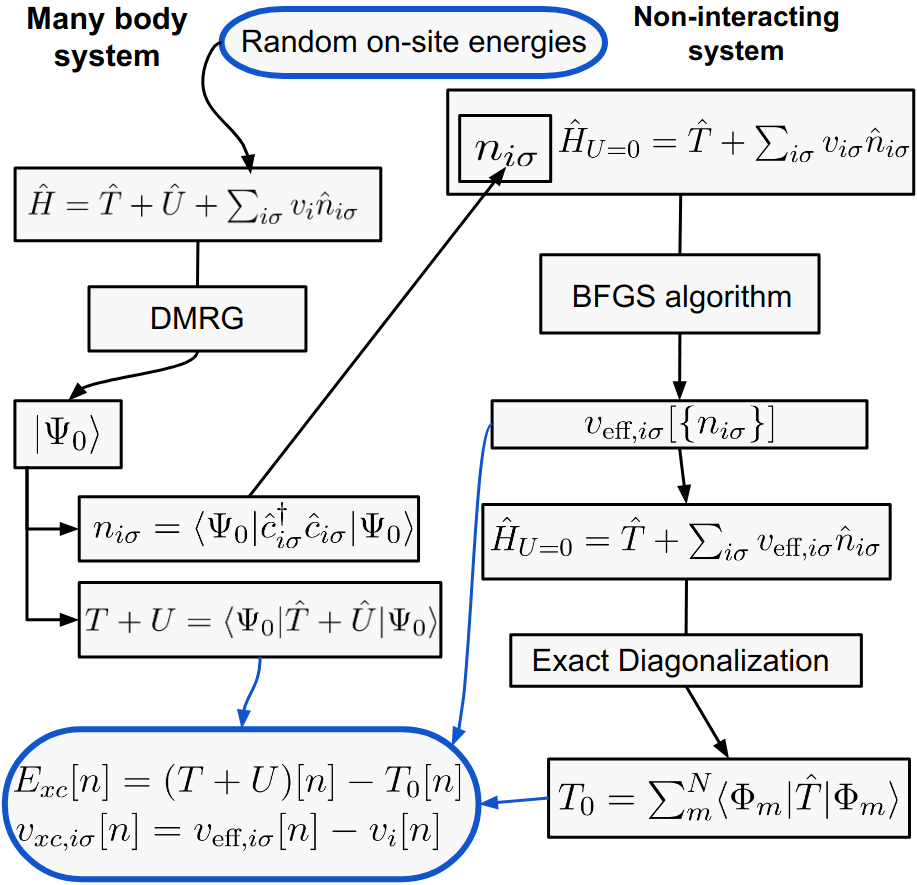}
    \caption{Workflow of the generation of the `exact' data. On the left-hand side we describe the
    DMRG pipeline. By starting with some random on-site-energy configuration, we use the DMRG algorithm 
    to find the ground-state wavefunction, density and energy. The construction of the `exact' KS
    potential is depicted on the right-hand side. The DMRG density is supplied to the 
    BFGS algorithm, along with the non-interacting Hamiltonian to find the effective potential that 
    reproduces the ground-state density. Finally, $E_{xc}[n]$ and $v_{xc,i}[n]$ are computed by 
    subtraction (blue box).}
    \label{fig:data_generation}
\end{figure}

As far as the test set is concerned, this is also generated by DMRG for a larger variety of lattice 
sizes, $L = 24, 30, 36, 42, 48, 54, 60$, with $n=1,000$ random external potentials taken from 
$\lambda^{(k)} = \sqrt{9k/(n-1)}$. In this case, we do not reject any DMRG results, not to create 
bias in the test set, namely, we want to avoid biasing the test set towards some particular class 
of disorder configurations. Since one of the target quantities is the KS exchange-correlation 
potential obtained as the functional derivative of the corresponding energy, we also need to generate 
test data for $v_\mathrm{xc}$. For this quantity we follow a well-established reverse engineering 
strategy, sometimes known as the inverse-Kohn-Sham method \cite{Shi2021}. This essentially consists 
of varying the effective potential until the targeted site occupations are obtained, namely those 
computed with DMRG. In practice, we minimize the Euclidean distance 
$\lVert n[v_{\text{eff}}] - n_\mathrm{DMRG} \rVert$ and the convergence is set by a threshold of 
10$^{-5}$. The same criterion is adopted for each lattice size investigated. In our case, we use the 
robust Broyden-Fletcher-Goldfarb-Shanno (BFGS) algorithm \cite{Broyden1970,Fletcher1970,Goldfarb1970,Shanno1970}, 
as implemented in SciPy \cite{scipy}, to descend onto the effective potential that reproduces the DMRG 
density. The workflow of the data generation is presented in Fig.~\ref{fig:data_generation}.
All of our data and some examples of our code are publicly available on GitHub \cite{git_repo}.

\subsection{\label{subsec3}Machine-learning the functional}
We have mentioned before, that our functional is constructed semi-local. This actually means 
that $E_\mathrm{xc}$ is obtained as the sum of site-dependent exchange-correlation energy densities
$e_\mathrm{xc}$, which in turn, depend on the occupation around that site within a given cutoff $a$, 
namely we have 
\begin{equation}
    E_{xc}^\mathrm{ML}[n] = \sum_i^{L+2a} e_\mathrm{xc}^\mathrm{ML} [\bar{n}_{i,a}]\;,
    \label{ehxc}
\end{equation}
where $\bar{n}_{i,a} = (n_{i-a}, n_{i-a+1}, .. , n_i, .. , n_{i+a-1}, n_{i+a})$, and $a$ is positive 
integer, which we call the non-locality parameter (cutoff radius). Note that the summation extends beyond 
indices associated to the actual lattice, since the non-locality assigns non-vanishing $\bar{n}_{i,a}$ 
vectors even when the index $i$ is outside the lattice boundaries. Then, the ML KS potential at site $i$ 
for spin $\sigma$, $v_{\mathrm{xc}, i\sigma}^\mathrm{ML}$, is computed by summing up
the partial derivative with respect to $n_{i\sigma}$ of all the energy-density terms that contain 
$n_{i\sigma}$ 
\begin{equation}
    v_{\mathrm{xc}, i\sigma}^\mathrm{ML}[n] = \frac{\partial}{\partial n_{i\sigma}} E_\mathrm{xc}^\mathrm{ML}[n] = 
    \frac{\partial}{\partial n_{i\sigma}}\sum_j e_\mathrm{xc}^\mathrm{ML} [\bar{n}_{j,a}]\:.
    \label{vhxc}
\end{equation}

The ML exchange-correlation energy is constructed as a fully connected neural network comprising five 
dense layers, each of 64 nodes, while the KS potential is obtained by the automatic differentiation as 
supplied by TensorFlow \cite{tensorflow2015-whitepaper}. We choose the exponential linear unit as an 
activation function to ensure that the functional derivatives are continuous. In each case, we 
train for 5,000 epochs with the Adam optimizer and a batch size of 32. One neural network is trained 
for each value of the non-locality parameter explored $a=1, 2, 3, 4$. The learning rate is decayed 
according to
\begin{equation}
    \lambda (j) = \lambda_0\Big(\frac{1}{30}\Big)^{j/5000}
    \label{decay}
\end{equation}
where $\lambda_0 = 3\cdot{10^{-4}}$ and $j$ is the epoch iteration. It is necessary to use zero-padding 
while computing the energy density near the boundary, as the non-local terms extend beyond it. In the 
context of LDFT, the use of zero padding can be understood as follows. Our finite system of $L$-sites 
with open boundary conditions can be considered to lie in the infinite chain with the on-site energies 
given by
\begin{equation}
    v_i =  \begin{cases} v_i & \text{if } 0 \leq i < L\\+\infty & \text{otherwise.}\end{cases}
\end{equation}
Corresponding to these on-site energies, the densities will be
\begin{equation}
    n_{i\sigma} = \begin{cases} n_{i\sigma}[\{v_i\}] & \text{if } 0 \leq i < L\\ 0              & \text{otherwise.} \end{cases}
\end{equation}

Data from larger systems can be understood to lie on the same infinite chain, the only difference 
being that some on-site energies at the boundary are brought down from infinity, allowing electrons 
to access those sites. Moreover, this compliments the fact that the $\hat{T}+\hat{U}$ component of the 
Hamiltonian is invariant, so that the same density-functional is valid for different numbers of lattice 
sites.

\section{Results \label{results}}
\subsection{\label{3a}Solving the Kohn-Sham Equations}
\begin{figure}[t]
    \centering
    \includegraphics[scale = 0.57]{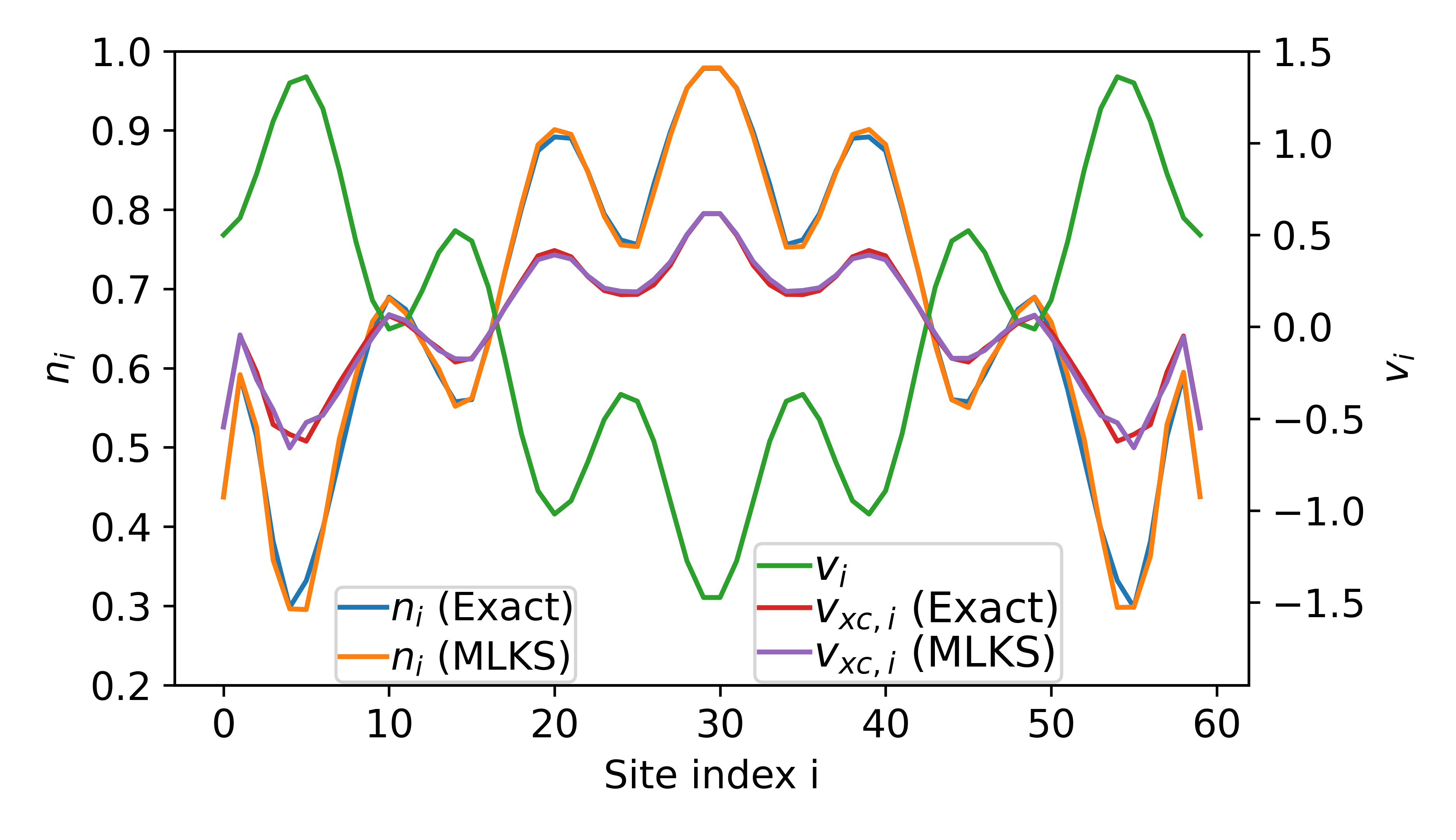}
    \caption{The ground-state site occupations (left-hand side scale) and the exchange-correlation potential (right-hand side scale)
    obtained from the MLKS scheme are plotted alongside the exact DMRG results (denoted as `exact'). For this visual comparison 
    we consider a $L=60$ system and an external potential (green line) of the form 
    $v_i = \cos\left(\frac{2\pi i}{L}\right) - \frac 12 \cos\left(\frac{12\pi i}{L}\right)$.
    }
    \label{Fig1}
\end{figure}

Let us start our analysis by giving a simple demonstration of a successful implementation of the 
trained functional and how to use it. In Fig.~\ref{Fig1} we present the site occupations and the 
converged KS exchange-correlation potential for a system of $L=60$ sites and an external potential 
of the form $v_i = \cos\left(\frac{2\pi i}{L}\right) - \frac 12 \cos\left(\frac{12\pi i}{L}\right)$. 
The functional in this case was trained over $L=18, 21, 24$ sites and here we compare the results 
from our ML KS (MLKS) scheme and the `exact' DMRG reference. The KS self-consistent cycle is 
performed with a simple linear-mixing scheme~\cite{Wagner2013}, in which the input set of site occupations 
at the $(k+1)$-th iteration, $n^{k+1}$, is constructed as $n^{k+1} = (1-\lambda)n^{k} + \lambda n^{\prime k}$, 
where $n^{\prime k}$ is the set of output occupations at iteration $k$. Here we set $\lambda$ to decay 
from $0.03$ to $0.005$ over the first 2,000 iterations and then to remain constant at $0.005$ 
should more iterations be required, and the convergence criterion is 
$\lVert n^{k} - n^{\prime k} \rVert < 10^{-5}$. As we can observe in Fig.~\ref{Fig1}, there is a good visual 
agreement between the MLKS-calculated quantities (occupations and potential) and the DMRG 
reference. In particular, $v_\mathrm{xc}$ appears to have the correct expected behaviour, namely it 
becomes more prominent in regions of high electron density, corresponding to minima of the external 
potential. 

\begin{figure}[t!]
    \centering
    \includegraphics[scale = 0.66]{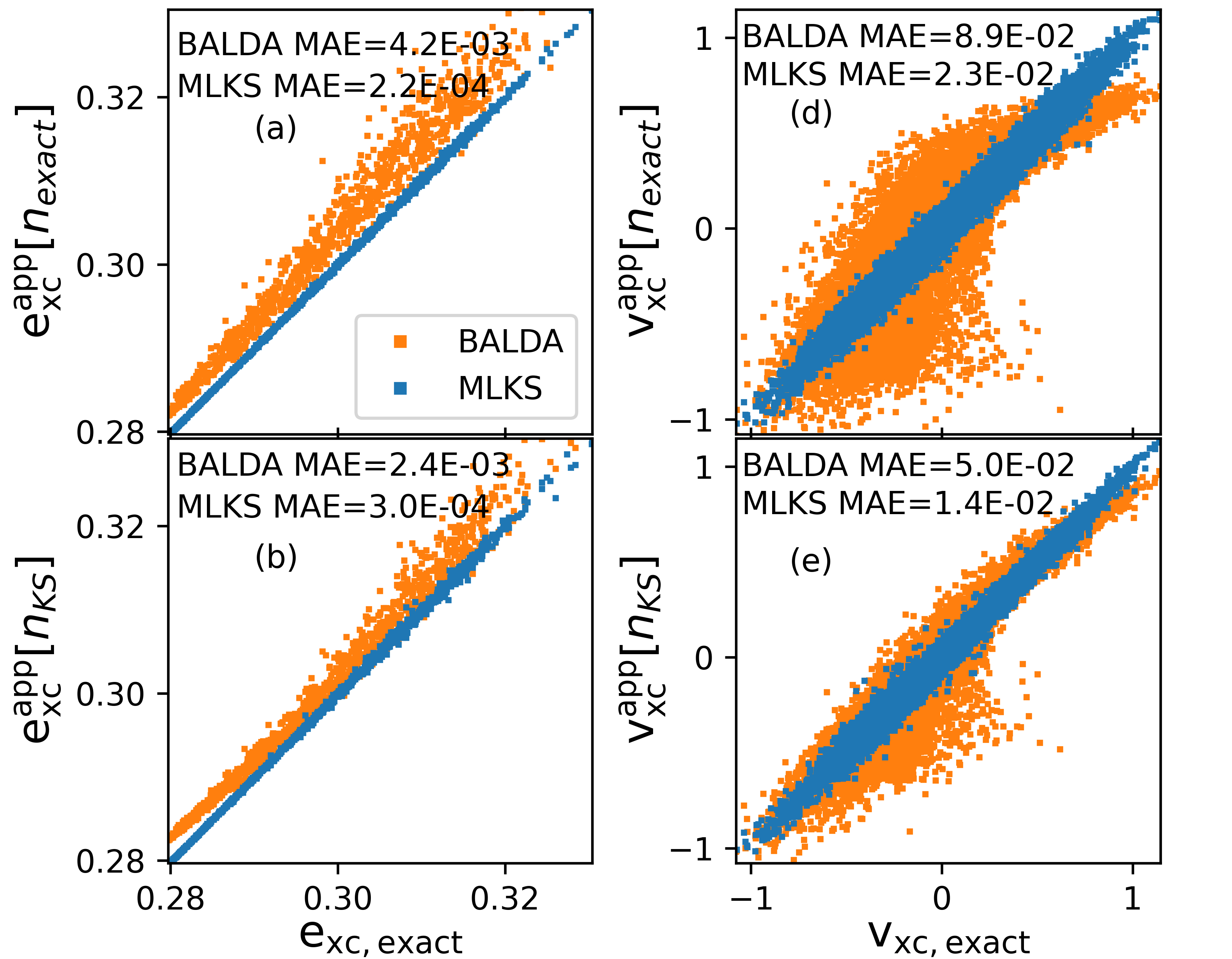}
    \includegraphics[scale = 0.625]{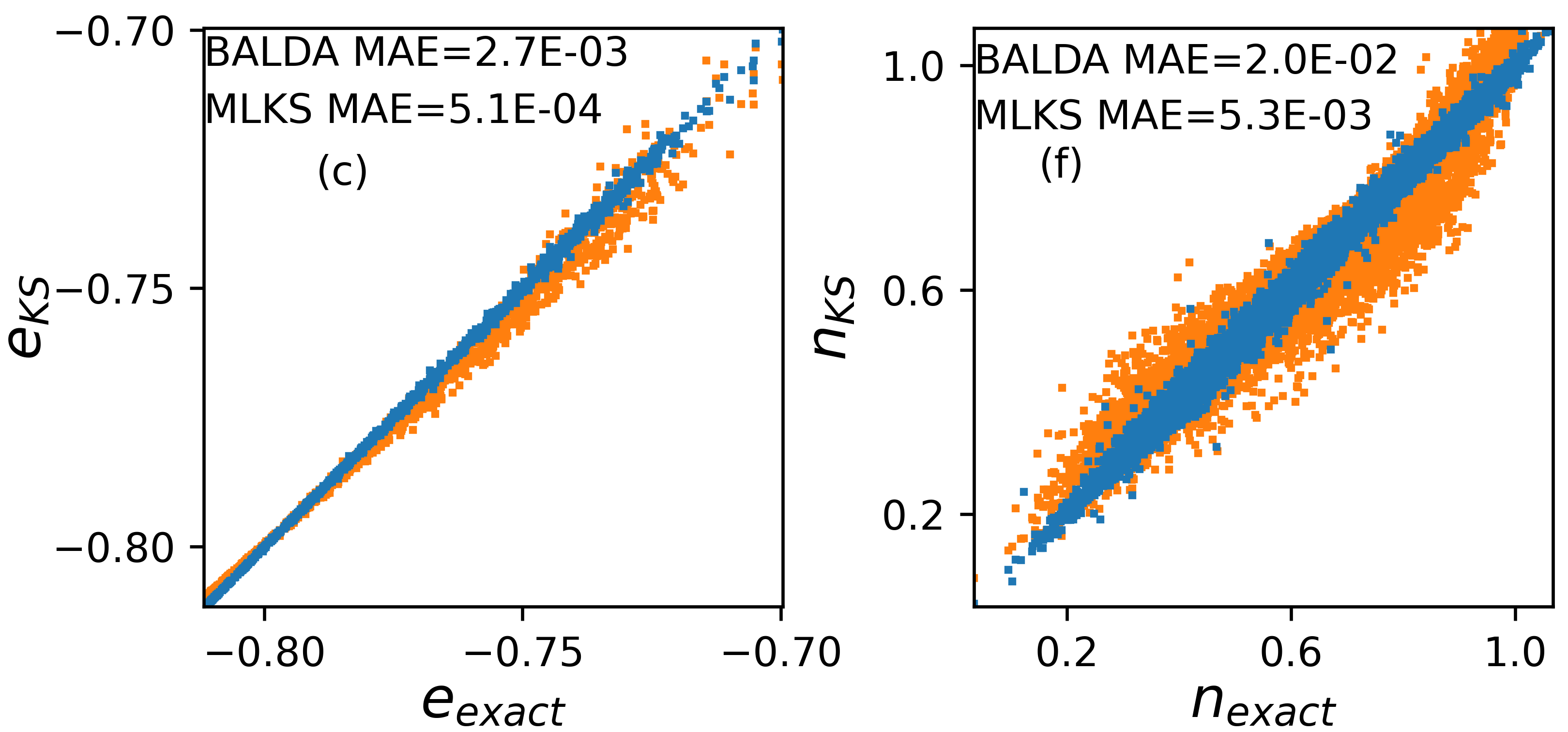}
    \includegraphics[scale = 0.625]{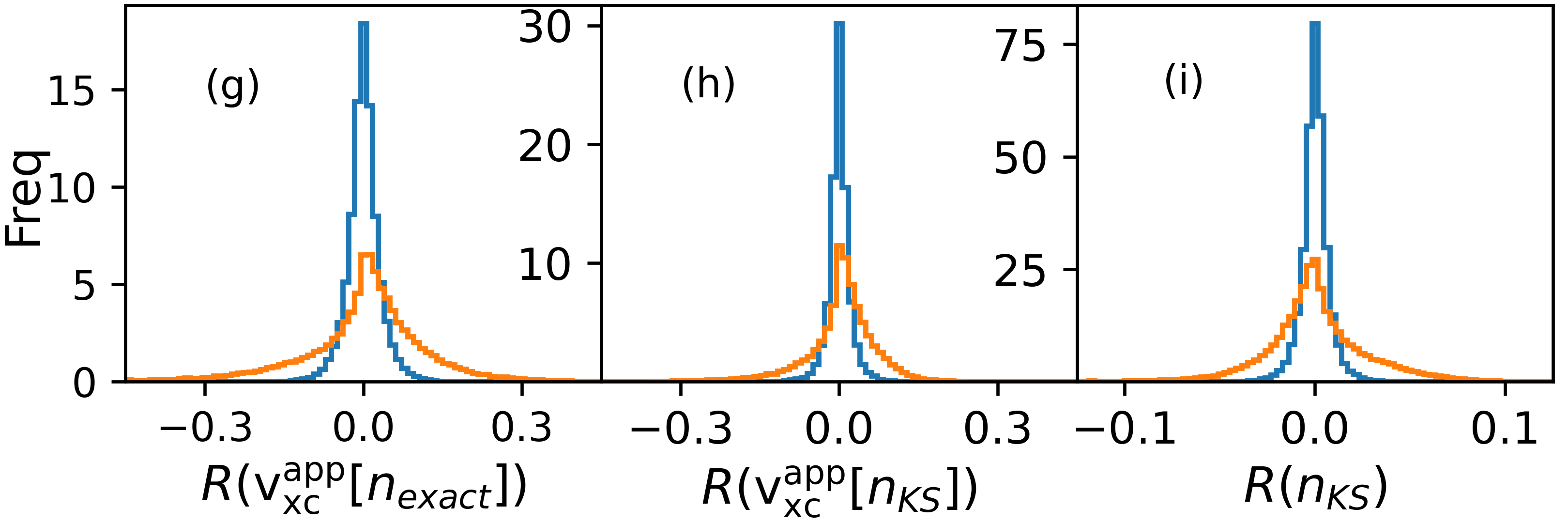}
    \caption{Analysis of the accuracy of the MLKS and BALDA functionals. For a system of $L=60$ sites, we 
    present the parity plots of both the ML exchange-correlation energy density, $e^\mathrm{app}_\mathrm{xc}$, 
    and the KS exchange-correlation potential, $v^\mathrm{app}_\mathrm{xc}$, computed either at the exact 
    (DMRG) occupations [panels (a) and (d), respectively] or at the converged KS occupations, $n_\mathrm{KS}$ 
    [panels (b) and (e), respectively]. The relevant approximate functional is denoted by `app', and it 
    corresponds to that indicated in the legend. In panels (c) and (f) we present the parity plots for the 
    KS total energy density and occupations, respectively. Panels (g) and (h) display histograms of the residuals 
    of $v_\mathrm{xc}^\mathrm{app}$ at the exact and converged KS densities, and panel (i) shows the densities. 
    Here, $R(x)$ denotes the residual quantity. For instance $R(n_\mathrm{KS}) = n_\mathrm{KS} - n_\mathrm{exact}$.
    Results are presented for the MLKS functional with a non-locality parameter $a=4$ and over a test set of 
    1,000 configurations. In the various panels we also report the MAE.
    }
    \label{nksb}
\end{figure}

It is important to note that when assessing the accuracy of the KS scheme at convergence one has to 
consider that the error in the KS potential itself adds to that of the ground-state site occupations, 
which are determined by the potential through the self-consistent cycle. It is then interesting to 
disentangle such errors by computing a number of additional quantities. In particular, in Fig~\ref{nksb} we 
compare both the exchange-correlation KS potential and energy density computed at either the exact DMRG 
density ($e_\mathrm{xc}^\mathrm{app}[n_\mathrm{exact}]$ and $v_\mathrm{xc}^\mathrm{app}[n_\mathrm{exact}]$, 
in panels (a) and (d), respectively) or the KS occupations ($e_\mathrm{xc}^\mathrm{app}[n_\mathrm{KS}]$ 
and $v_\mathrm{xc}^\mathrm{app}[n_\mathrm{KS}]$, in panels (b) and (e), respectively), with the corresponding 
exact (DMRG) quantities. Here, `app' denotes the relevant approximate functional. Namely, our MLKS functional, 
or the fully-numerical implementation of the Bethe-ansatz LDA (BALDA) \cite{SGN1995, Shiba1972, Knizia2012, Chong2020, balda-repo}.
The BALDA is perhaps the most developed non-ML approximation to the exchange-correlation functional for the 
Hubbard model. It can be used both in a orbital-free DFT mode or within a KS scheme. Here it will represent a point
of comparison for our developed functional. For the results of Fig.~\ref{nksb} we consider a test system of 60 sites 
and the various parity plots are constructed over 1,000 disorder configurations.

In the case of MLKS, it appears that the error accumulated over the KS self-consistent cycle is only minor. 
In fact, we find a mean absolute (MAE) error over the exchange-correlation energy density of 
$2.2\times 10^{-4}$, when this is computed at the exact DMRG occupations, a MAE that grows to only 
$3\times 10^{-4}$ with the converged KS occupations. Note that $e_\mathrm{xc}^\mathrm{ML}$ averages at 
around 0.3 for this choice of disorder and occupation, meaning that the relative error on the energy 
density is of the order of 0.1\%. Interestingly, the results for the KS exchange-correlation potential 
returns a situation where the MAE is lower, even if marginally when $v_\mathrm{xc}^\mathrm{ML}$
is computed at the KS occupations rather than at the exact ones. This suggests that the functional 
derivative defining the KS potential may slightly shift the ground-state occupations from their exact 
values. In any case, an overall assessment of the model constructed can be obtained from panels (c) and 
(f) of Fig.~\ref{nksb}, where we show the parity plot for the total energy density, $e_\mathrm{MLKS}=F_U/L$ 
and the KS occupations, respectively. These give us relative MAEs of 0.07\% ($e_\mathrm{MLKS}$) and
1\% ($n$), pointing to an extremely accurate model, whose main error accumulates in the self-consistent 
occupations. 

Unlike the MLKS scheme, the BALDA results show some systematic error. Namely, $e_\mathrm{xc}$ appears to 
be over-approximated, while $v_\mathrm{xc}$ has some curvature relative to the parity line. Visually, the 
latter error [see Fig.~\ref{nksb} panels (d) through (f)] is inflated by the high number of data points 
on the plot, 60,000, where the majority of the points lie relatively close to the parity line, although 
they are not visible. Therefore, for a better visualization, 
we plot a histogram of the residuals in panels (g) through (i). Similarly to the case of MLKS, but to a 
much greater extent, the BALDA results improve dramatically, when computed at the converged KS density 
instead of the exact one. In fact, the errors on $e_\mathrm{xc}$ and $v_\mathrm{xc}$ drop roughly by a 
factor two. Furthermore, in panel (c) the systematic over-approximation of $e_\mathrm{xc}$ appears to 
be cancelled out by an under-approximation of the non-interacting kinetic energy, giving a $e_\mathrm{KS}$ 
in a satisfactory agreement with the exact data. 

All considered, the MLKS functional appears to be in much better agreement with the exact functional
than the well-known BALDA. The evidence for this is (i) the consistently lower values of MAE for each 
quantity, (ii) the absence of systematic errors in the parity plots and (iii) the higher degree of 
consistency between the results calculated at the exact and KS densities. In Appendix~\ref{a=0_parity_plots_section}, 
we repeat the same test, using the fully local $a=0$ ML functional and discuss how that compares to 
the BALDA.

\subsection{System size and non-locality parameter}\label{a_dependence}
\begin{figure}
    \centering
    \includegraphics[scale = 0.7]{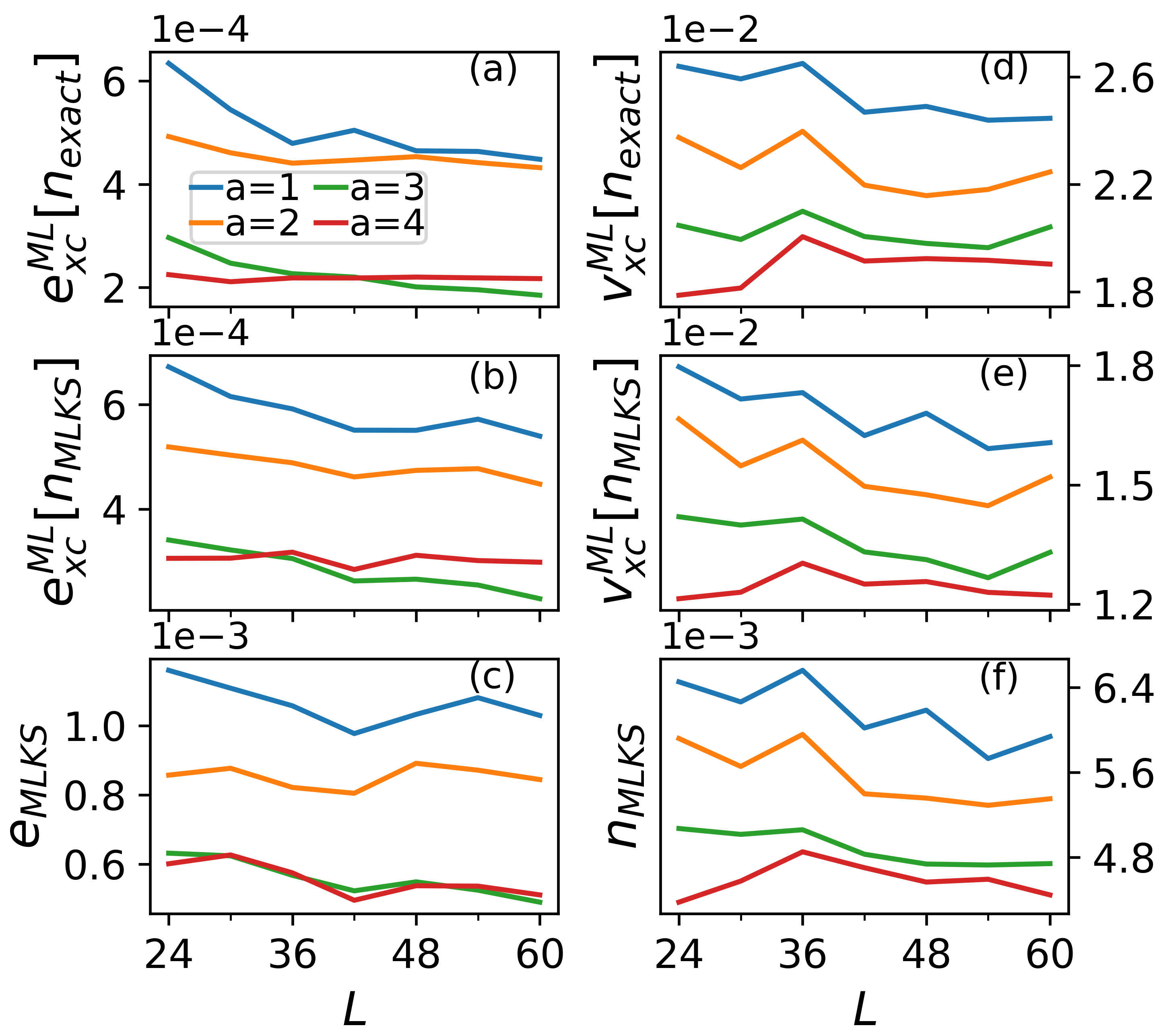}
    \caption{The MAEs, computed over 1,000 disorder configurations for each system, as a function of the system size, $L$. Data
    are presented for the ML exchange-correlation energy density, $e^\mathrm{ML}_\mathrm{xc}$, and the KS exchange-correlation 
    potential, $v^\mathrm{ML}_\mathrm{xc}$, computed either at the exact (DMRG) occupations [panels (a) and (d), respectively]
    or at the converged KS occupations [panels (b) and (e), respectively]. In panels (c) and (f) we present the MAE for the ML KS 
    total energy density and occupations, respectively. In all cases, we show results for different  values of the non-locality parameter
    $a$.}
    \label{a_dep}
\end{figure}

In order to have a better insight into the ML functional construction, it is useful to investigate its 
non-local properties. The effects of the functional built-in non-locality are analysed by looking at lattice 
sizes ranging from 24 to 60, and by computing the MAEs for different values of the non-locality, $a=1, 2, 3, 4$. 
The results ensemble-averaged over 1,000 disorder configurations are summarized in Fig~\ref{a_dep},
where we present the MAEs of $e^\mathrm{ML}_\mathrm{xc}[n_\mathrm{exact}]$ and 
$e^\mathrm{ML}_\mathrm{xc}[n_\mathrm{MLKS}]$ in panels (a) and (b), $v^\mathrm{ML}_\mathrm{xc}[n_\mathrm{exact}]$ 
and $v^\mathrm{ML}_\mathrm{xc}[n_\mathrm{MLKS}]$ in panels (d) and (e), and again the 
MAEs of $e_\mathrm{MLKS}=F_U/L$ and the KS occupations in panels (c) and (f). A number of trends 
are clearly visible. 

Firstly, one may notice that the functionals get progressively more accurate with increasing their 
non-locality $a$, in particular when going from $a=1$ to $a=3$, after which the performance improvement 
seems to saturate. This is particularly true for the energy density $e^\mathrm{ML}_\mathrm{xc}$, while 
further improvements at $a=4$ are found for the potential $v^\mathrm{ML}_\mathrm{xc}$. Such numerical
result indicates that the non-local correlation of the Hubbard model at the chosen filling factor is 
well captured at $a\simeq 3$. It is also interesting to note that the smaller MAE for 
$v^\mathrm{ML}_\mathrm{xc}$ is computed at the MLKS occupations and not at the exact ones, 
$v^\mathrm{ML}_\mathrm{xc}[n_\mathrm{exact}]$. This behaviour was observed before for $a=4$ and it is 
found here for all systems regardless of the non-locality. 

\begin{figure}
    \centering
    \includegraphics[scale = 0.75]{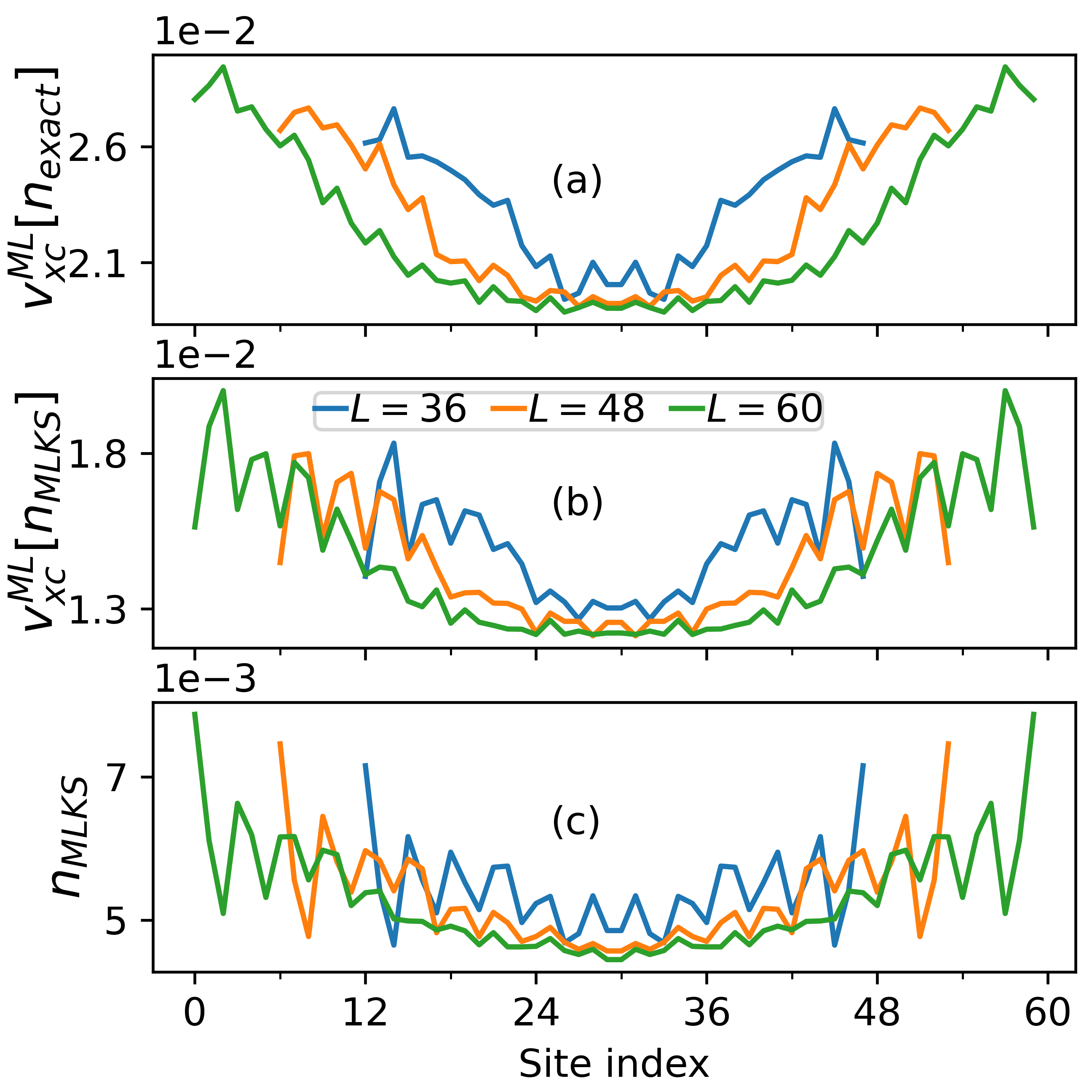}
    \caption{Site-resolved MAE, computed over 1,000 disorder configurations for three lattice sizes, 
    $L = 36, 48, 60$ and non-locality, $a=4$. The (a) MAE for $v^\mathrm{ML}_\mathrm{xc}$ at exact 
    site occupations, $n_\mathrm{exact}$, and (b) at the MLKS ones, $n_\mathrm{MLKS}$. In panel (c) 
    we show the MAE of the MLKS site occupations (see text for details).}
    \label{boundariesplot}
\end{figure}
The second noteworthy trend is that, in general, the MAE appears to decrease with increasing the lattice 
size. To investigate whether this is a boundary effect, we look into the site dependence of the MAEs as function 
of the lattice size. In Fig.~\ref{boundariesplot} we show the site-resolved MAE of the KS exchange-correlation 
potential, computed at both the MLKS and exact occupations, and the final occupations, as a function of the 
lattice position. This is shown for three different lattices, $L=36, 48$ and $60$, and the MAEs are averaged 
over 1,000 different disorder configurations. We find that, on average, the MAE at the edges of the lattice 
is about 25\% larger than in the middle, suggesting that the systematic trend in the MAE decrease with system 
size may be due to a decreasing edge-to-bulk ratio. In contrast, in the middle of the lattice, all quantities 
converge to a similar MAE, regardless of the lattice size. The loss of accuracy at the lattice edges may be 
attributed to an imbalance in the training data. In fact, the edges are characterised by 
$\bar{n}_{i,a} = (n_{i-a}, n_{i-a+1}, .. , n_i, .. , n_{i+a-1}, n_{i+a})$ vectors presenting $n_j=0$ for 
those sites lying outside the lattice. Vectors of this type, however, are less abundant in our training set. 
In fact, for a lattice of $L$ sites and a non-local parameter $a$, there are $4a$ sites with $\bar{n}_{i,a}$ 
presenting at least one vanishing occupation and $L-2a$ sites with all non-vanishing occupations.

In any case, our results return us a rather accurate exchange-correlation functional, which
can be used to derive a KS potential via functional derivative. Already with a modest degree of 
non-locality, this allows us to compute energies and occupations close to those obtained with DMRG, 
with the accuracy improving as the system size grows. 

\subsection{Extrapolation to different filling factors \label{extrapolation_section}}
The universal DFT functional is independent of the total number of particles in the system. However,
there is no guarantee that this property is transferred to approximated machine-learning formulations.
Namely, although one can always extend a semi-local functional to filling factors outside the training
range, it remains the question of whether the functional is actually accurate. This question naturally 
made us curious to test how well the trained functionals extrapolate to filling factors away from 2/3.
By using both the DMRG and the MLKS scheme, we now map the density, energy and exchange-correlation potential
of 128 random configurations of a $L=36$ lattice and different electron numbers, 
$N_\mathrm{e} = 16,18,...,32$ (the training was performed for $N_\mathrm{e} = 24$). In this test, we
use the same $\lambda^{(k)} = \sqrt{9k/(n-1)}$ distribution described in Sec.~\ref{data_generation_section} 
to generate the disorder configurations, and examine how the average error depends on the number of electrons. 

\begin{figure}[H]
    \centering
    \includegraphics[scale = 0.73]{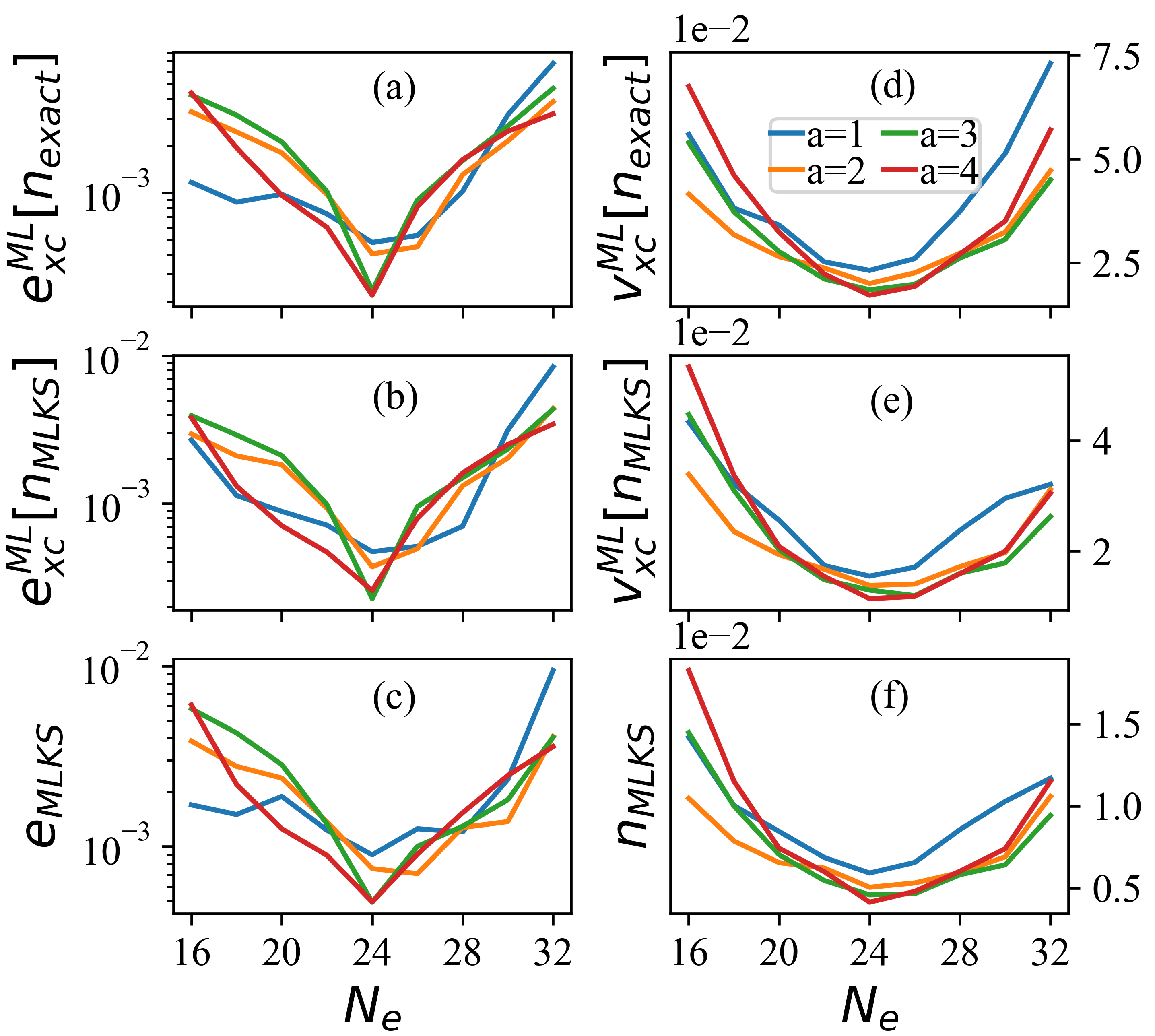}
    \caption{MAE obtained over 128 disordered configurations for the exchange-correlation energy
    density and KS potential of a system of 36 sites and different number of electrons, $N_\mathrm{e}$.
    The model has been trained at $N_\mathrm{e} = 24$, where the highest accuracy is reached. The 
    various quantities have been defined in the main text.}
    \label{mae_vs_filling}
\end{figure}

The results are given in Fig.~\ref{mae_vs_filling}. Unsurprisingly, the error minimizes at 
$N_\mathrm{e}/L=\frac{2}{3}$, namely at the same filling of the training set, where all energies densities
shown in panels (a) through (c) take on minima of the order of $10^{-4}$ and have similar scaling away 
from this point. For instance, the converged Kohn-Sham energy density [panel (c), $a=4$] has a MAE of 
$2\times10^{-3}$, $5\times10^{-4}$, $2\times10^{-3}$ at $N_\mathrm{e} = 18, 24$ and 30, respectively. 
In the case of the exchange-correlation potential and similarly for the occupation, the MAE varies 
more slowly against the number of electrons. In panel (f), the $a=4$ curve has a MAE of $6\times 10^{-3}$, 
$2\times 10^{-3}$, $4\times 10^{-3}$ at $N_\mathrm{e} = 18, 24, 30$, respectively. Finally,
it is worth noting that the extrapolation of the functional (both energy density and potential)
does not seem to have a strong dependence on the degree of non-locality.

The success of the model's extrapolation is likely due to the inhomogeneous nature of the training 
data, combined with the semi-local architecture of the functional. This gives the model exposure to 
density regions with occupations both higher and lower than 2/3 filling. It can be expected that this 
extrapolation (the functional accuracy) will improve as the system size is increased, provided that 
only a few electrons are added to the system, since $(N_e \pm \delta N_e)/ L \approx N_e/L$ for large 
$L$ (namely for variations close to the training range). The ability of the model to extrapolate is a 
useful property, as it allows one to perform tests, which require $N_\mathrm{e}$ to be varied. For 
instance, one could compute the band gap, or the inverse compressibility~\cite{Mila1993, Xianlong2006}. 
In particular, in the next section~\ref{piecewise_linearity_section}, we test whether or not the 
KS energy is piecewise linear.

\subsection{\label{piecewise_linearity_section}Piecewise Linearity of the KS energy}
Conventional DFT can be extended to fractional numbers of electrons by using a zero-temperature
ensemble of integer electron states. Within this framework, it can be shown that the exact DFT total 
energy computed at integer electron-number values connects linearly~\cite{Perdew1982}, namely the 
$E-N_\mathrm{e}$ curve is piecewise linear. This property formally writes 
\begin{equation}
    \label{piecewise_linearity}
    E(N) = (1-x)E(N_0) + xE(N_0+1)\:,
\end{equation}
where $N=N_0+x$, $N_0$ is an integer and $x$ is taken in the interval $[0,1]$. The piecewise linearity, however, is 
not satisfied by semi-local density functionals~\cite{MoriSanchez2006, Ruzsinszky2007, Vydrov2007, Cohen2008, Stein2012, Haunschild2012, Cohen2012, Kraisler2013}. For instance, in the local density approximation such curve is approximately 
parabolic around the neutrality point. The shortfall is widely attributed to the `delocalization error' or the 
related `self-interaction error', which gives rise to a set of hallmark DFT deficiencies: an inaccurate 
description of charge transfer and charge distribution in separated systems, a systematic under-estimation 
of the band gaps, and generally over delocalized orbitals~\cite{Armiento2013}. 

If Eq.~(\ref{piecewise_linearity}) holds, then the highest occupied Kohn-Sham orbital at $N=N_0$,
$\epsilon_\mathrm{H}^{N_0}$, is related to the slope of $E(N)$ as
\begin{equation}
    \label{ionization_potential}
    \epsilon_\mathrm{H}^{N_0} = \frac{\partial E}{\partial N} = E(N_0) - E(N_0-1)\:,
\end{equation}
so that $-\epsilon_\mathrm{H}^{N_0}$ equals the ionization potential, $IP$. The same argument can then be applied 
to the case of $N_0+1$ electrons, where one finds the electron affinity, $EA$, of the $N_0$ electron system. Thus, one 
can conclude that the fundamental gap of a molecule containing $N_0$ electrons is simply
\begin{equation}\label{gap}
    \Delta^{N_0} = IP-EA = \lim _{x \rightarrow 0}\left\{\left.\frac{\partial E}{\partial N}\right|_{N_0+x}-\left.\frac{\partial E}{\partial N}\right|_{N_0 - x}\right\}\:.
\end{equation}
In general, $\Delta^{N_0}$ is not the difference between the lowest-unoccupied and the highest-occupied Kohn-Sham orbital
(the Kohn-Sham gap), $\Delta^{N_0}_\mathrm{KS}=\epsilon_\mathrm{L}^{N_0} - \epsilon_\mathrm{H}^{N_0}$, a fact attributed 
to the lack of derivative discontinuity of the exchange-correlation energy. 

\begin{figure}[H]
    \centering
    \includegraphics[scale = 0.8]{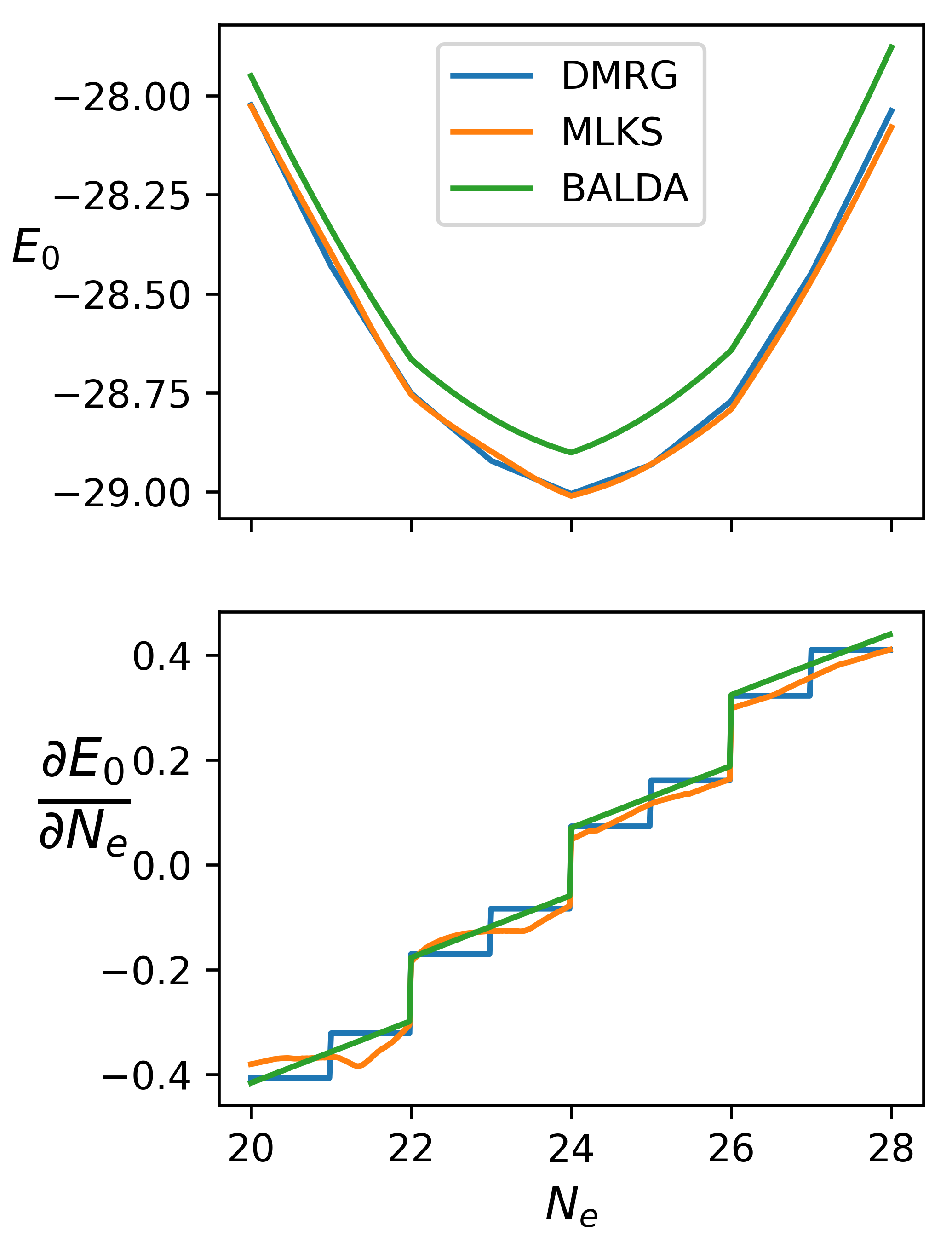}
    \caption{Total energy against the number of electrons (top panel) and its derivative (bottom panel)
    for different functionals. Results are presented for a homogeneous $L=36$ lattice. The blue curves 
    (DMRG) are computed by DMRG at integer electron numbers and interpolated linearly using 
    Eq.~(\ref{piecewise_linearity}). The BALDA and MLKS curves are computed by solving the Kohn-Sham 
    equations at fractional occupation. In the case of MLKS we have used the functional with $a=4$.}
    \label{pieccewise_linearity_plot}
\end{figure}

We now test both the MLKS functional (we focus here only on the $a=4$ case) and the BALDA against this behaviour 
for a homogeneous $L=36$ lattice, which can be viewed as a large molecule. In the upper panel of Fig.~\ref{pieccewise_linearity_plot}
we present the total energy as a function of the number of electrons, while the lower panel shows its derivative. The exact
limit is established by the DMRG results. In this case we have access only to integer values of the electron number and 
the energy, by construction, is piecewise linear at any integer. In contrast, the total energy of both the MLKS functional 
and the BALDA changes slope only at even electron numbers, while in between those the $E(N_\mathrm{e})$ curve is
convex. This is because the MLKS functional has been trained only for the $N_{\uparrow} = N_{\downarrow}$ case and
we do not have access to broken symmetry solutions. Note that in standard DFT the exact functional should satisfy the 
flat-plane condition with respect to the magnetization~\cite{Yang2000,Cohen2008b}, so that one does not need a broken 
symmetry solution to satisfy the derivative discontinuity at odd electron count. In our case the MLKS functional
has never experienced any configurations with an odd number of electrons, so that the highest occupied Kohn-Sham orbital 
is continuously filled until it accommodates two electrons under the constrain $N_{\uparrow} = N_{\downarrow}$.
This gives rise to a linear dependence of the energy derivative (see lower panel of Fig.~\ref{pieccewise_linearity_plot}). Such 
dependence is only partially satisfied for $N_0<24$, probably due to some numerical error.
%


\begin{table*}
\caption{System gap computed at even-integer electron numbers, $N_0$. In the table we report the DMRG gap, 
$\Delta_\mathrm{DMRG}^{N_0}$, the LDFT one computed from Eq.~(\ref{gap}), $\Delta_\mathrm{app}^{N_0}$,
and the Kohn-Sham gap, $\Delta^{N_0}_\mathrm{KS}$. In addition we report the corresponding percentage 
errors, $\delta_\mathrm{app}^{N_0}$ and $\delta_\mathrm{KS}^{N_0}$, respectively. Data correspond to those 
presented in Fig.~\ref{pieccewise_linearity_plot} for a homogeneous $L=36$ lattice.}
\label{dd_table}
\begin{ruledtabular}
\begin{tabular}{cccccccccc}
 &\multicolumn{1}{c}{DMRG}&\multicolumn{4}{c}{MLKS}&\multicolumn{4}{c}{BALDA}\\
 $N_0$&$\Delta_\mathrm{DMRG}^{N_0}$&$\Delta_\mathrm{app}^{N_0}$&$\delta_\mathrm{app}^{N_0}$ (\%)&$\Delta^{N_0}_\mathrm{KS}$ & $\delta_\mathrm{KS}^{N_0}$ (\%)&$\Delta_\mathrm{app}^{N_0}$&$\delta_\mathrm{app}^{N_0} (\%)$&$\Delta^{N_0}_\mathrm{KS}$ & $\delta_\mathrm{KS}^{N_0}$ (\%)\\ \hline
22&	0.151&	0.121&	20.3&	0.127&	16.1&	0.122&	19.1&	0.127&	15.8\\
24&	0.157&	0.127&	19.4&	0.135&	13.7&	0.130&	17.4&	0.138&	12.0\\
26&	0.162&	0.135&	16.3&	0.144&	10.8&	0.136&	15.5&	0.142&	12.0\\
28&	0.164&	0.141&	13.9&	0.148&	10.1&	0.142&	13.7&	0.149&	9.1\\
\end{tabular}
\end{ruledtabular}
\end{table*}

The linearity of $\frac{\partial E}{\partial M}$ between even integers implies that in that interval the Kohn-Sham energy is
parabolic with the electron filling. This means that the Hartree contribution to the total energy, 
$E_\mathrm{H} = \frac{U}{4}\sum_i n_i^2$, is not compensated by the curvature of the exchange-correlation energy
(here we refer to the conventional definition of exchange-correlation energy, since the exchange-correlation energy
of the MLKS functional contains the Hartree term). The persistence of this error through the BALDA and the MLKS
functional is thus inherent to (semi-) local functionals, as is the case in \textit{ab-inito} DFT. 

To conclude this section we verify the accuracy of the constructed functional at describing the system gap, as defined
in Eq.~(\ref{gap}). In this case we have access to the exact DMRG result, $\Delta_\mathrm{DMRG}^{N_0}$, and to
the LDFT values, $\Delta_\mathrm{app}^{N_0}$. The comparison is carried out only at even integers where 
the LDFT gaps are defined, with the results presented in Table~\ref{dd_table}. In particular, together with 
$\Delta_\mathrm{app}^{N_0}$ we also report the computed Kohn-Sham gap, $\Delta^{N_0}_\mathrm{KS}$, and
the corresponding percentage errors, 
$\delta_\alpha^{N_0}(\Delta_\mathrm{DMRG}^{N_0}-\Delta^{N_0}_\alpha)/\Delta_\mathrm{DMRG}^{N_0}$,
with $\alpha=$ `app' and `KS'.

From the table it emerges clear that $\Delta^{N_0}$ computed at the LDFT level remains systematically smaller
than its corresponding exact DMRG result. In particular, we find little difference between the BALDA and our
MLKS functional, both returning an underestimation of the gaps in between 13\% and 20\%. Importantly,
the Kohn-Sham gap, $\Delta^{N_0}_\mathrm{KS}$, is quite close to $\Delta_\mathrm{app}^{N_0}$, so that
the exchange and correlation energy has no derivative discontinuity, both in the case of MLKS and the BALDA.
Note that this has not to be confused with the discontinuity that the BALDA presents at half-filling~\cite{Lima2002}, where 
the functional fits the Mott transition of the uniform infinite 1D Hubbard model.

\subsection{Response to an electric field \label{polarizability_section}}

In order to demonstrate the use of our ML KS scheme we explore the response 
of the one-dimensional Hubbard model to an external, constant electric field 
$\mathcal{E}$. The response is quantified by the polarizability, $\alpha$, 
defined as the derivative of the induced electrical dipole moment, $P(\mathcal{E})$,
with respect to the electric field, in the limit $\mathcal{E}\rightarrow0$. 
The presence of the electric field adds a linear on-site potential 
$v_{{\cal E}i}=e{\cal E}(i-\bar{x})$, which introduces a new term to the Hamiltonian,
\begin{equation}\label{H-Ele}
\hat{H}_{\mathcal{E}}=\sum_\sigma\sum_{i=1}^L v_{{\cal E}i} \hat{c}_{i\sigma}^{\dagger} \hat{c}_{i\sigma}=e \mathcal{E} \sum_\sigma\sum_{i=1}^L(i-\bar{x}) \hat{c}_{i\sigma}^{\dagger} \hat{c}_{i\sigma}\:,
\end{equation}
where $\bar{x}$ is the central position in the lattice. The dipole moment can 
then be computed as the expectation value of the dipole operator,
\begin{equation}\label{P(E)-MB}
P(\mathcal{E})=e\left\langle\Psi_0(\mathcal{E})\left|\sum_\sigma\sum_{i=1}^L(i-\vec{x}) \hat{c}_{i\sigma}^{\dagger} \hat{c}_{i\sigma}\right| \Psi_0(\mathcal{E})\right\rangle\:,
\end{equation}
where $|\Psi_0(\mathcal{E})\rangle$ is the many-body ground-state wave-function 
for the given electric field, $\cal E$. The same procedure can then be followed 
by using our ML KS scheme, by solving the self-consistent KS equations in the 
presence of the electric field term of Eq.~(\ref{H-Ele}), and then by computing
\begin{equation}\label{P(E)-KS}
P(\mathcal{E})^\mathrm{KS}=e\sum_\sigma\sum_{i=1}^L(i-\vec{x}) n^\mathrm{MLKS}_{i\sigma}\:.
\end{equation}
The polarizability is then obtained as the numerical derivative of $P(\mathcal{E})$ 
at $\mathcal{E} = 0$,
\begin{equation}
    \label{alpha}
    \alpha =  \left.\frac{dP(\mathcal{E})}{d\mathcal{E}}\right|_{\mathcal{E}=0}\:.
\end{equation}

\subsubsection{Polarizability scaling}
\begin{figure}[H]
    \centering
    \includegraphics[scale = 0.085]{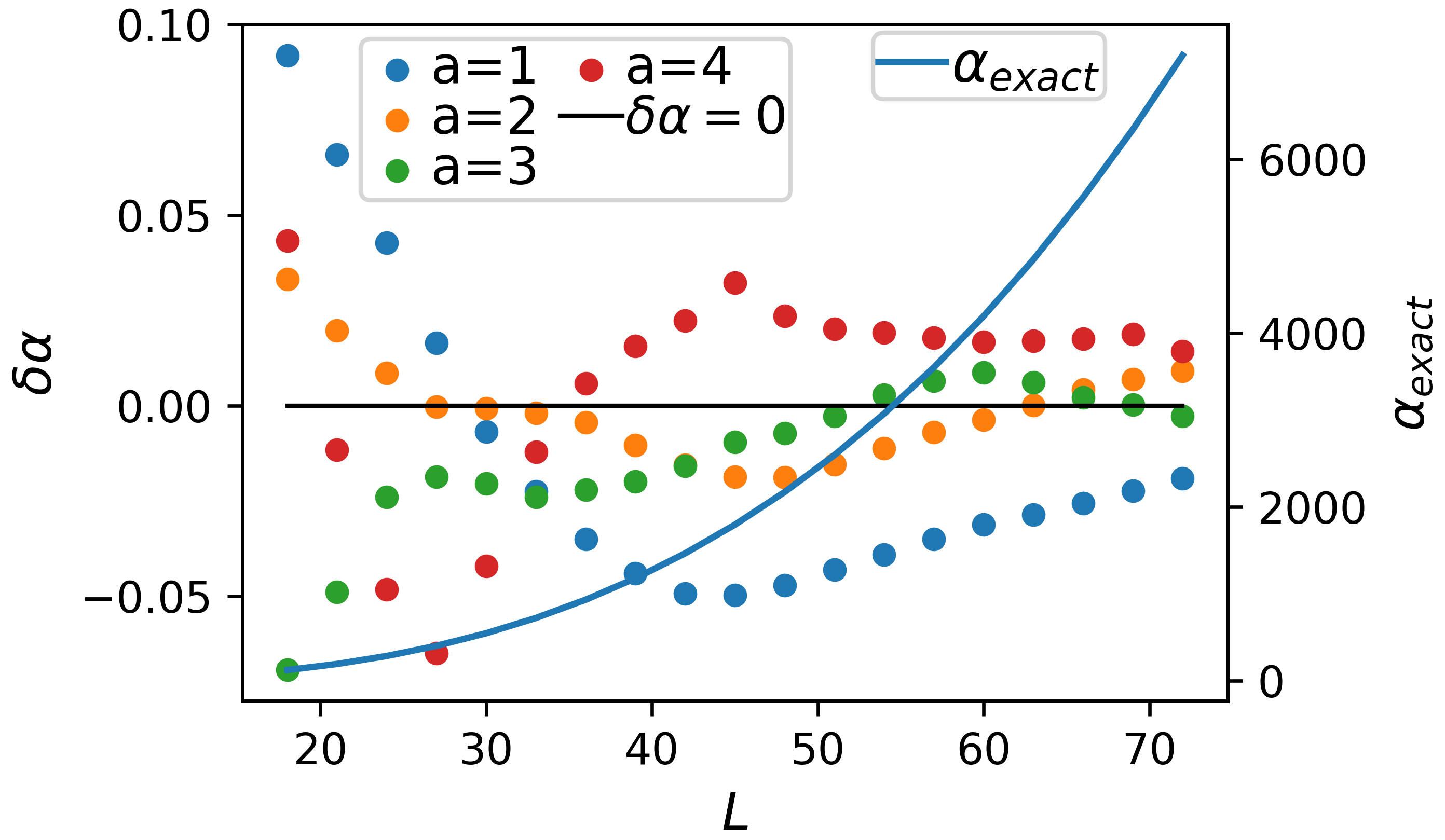}
    \caption{Polarizibility, $\alpha$, of the 1D Hubbard model at one-third-filling, 
    plotted against the number of sites, $L$, of a linear open-ended chain. Here 
    $\alpha_\mathrm{exact}$ is computed with DMRG (blue line - right-hand-side scale), 
    while the symbols show the fractional error, 
    $\delta \alpha = (\alpha_\mathrm{MLKS} - \alpha_\mathrm{exact})/\alpha_\mathrm{exact}$, 
    of the MLKS-calculated polarizability scheme relative to the exact results 
    (left-hand side $y$-axis). The calculation is repeated for functionals having different non-locality parameter, $a$.}
    \label{pscaling}
\end{figure}
Our computed polarizability, $\alpha$, as a function of the system length, 
$L$, is presented in Fig.~\ref{pscaling} for homogeneous lattices of size
$L=18, 21, 24, ..., 72$. In the figure we present the exact DMRG results together 
with the errors of our ML KS scheme for models constructed with different 
non-locality parameters. Clearly, there is an excellent agreement between the 
DMRG and MLKS results across the entire range of lengths, $L$. The agreement
is particularly good for long chains and non-local functionals, and remains
always below 5\%, except for the small lattices and the $a=1$ case.
This reflects the already noted improvement of the exchange-correlation 
functional for large systems arising from the reduced relevance of the edge 
occupations. Furthermore, it is clear that the non-locality improves the 
quality of the ML KS results, although there is little quantitative 
difference between different degrees of non-locality.

\begin{figure}[H]
    \centering
    \includegraphics[scale=0.7]{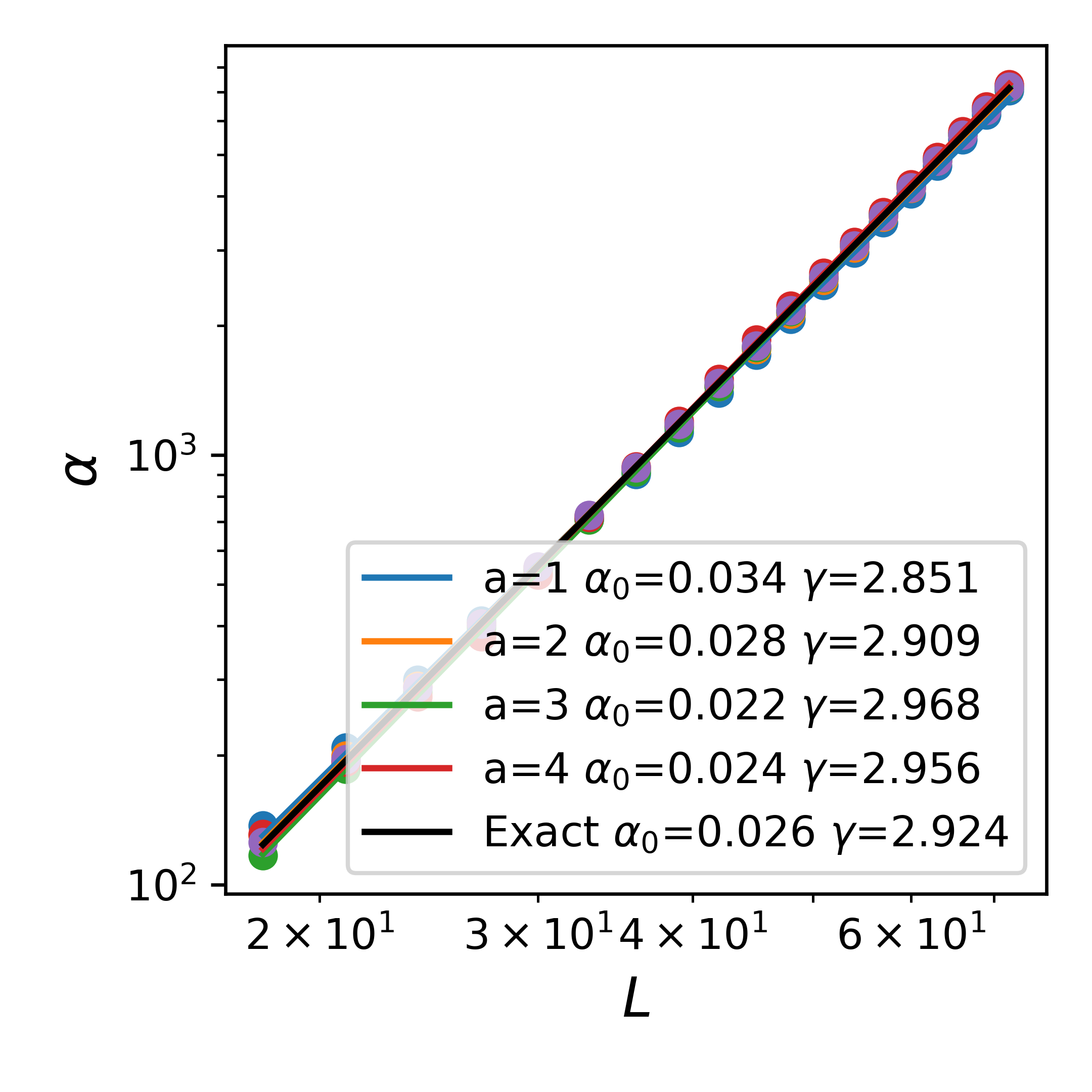}
    \caption{Polarizability, $\alpha$ versus lattice size, $L$, plotted on a 
    log-log scale. The legend distinguishes between the various MLKS functionals 
    and the exact results. Also given in the legend are the corresponding values 
    found from fitting the function $\alpha(U; L)=\alpha_0L^\gamma$.}
    \label{pol_fit}
\end{figure}

The polarizability as a function of system size follows a 
power-law \cite{Akande2010} behaviour
$$
\alpha(U; L)=\alpha_0L^\gamma\:.
$$
A numerical fit to the above scaling law [see Fig.~\ref{pol_fit}] yields us exponents 
$\gamma=2.851, 2.909,2.968, 2.956$ for non-locality $a=1, 2, 3$ and 4, respectively.
This is very close to $\gamma=3$ expected for free electrons in 1D \cite{Rojo1993} 
and in line with results obtained with Bethe-ansatz LDFT results~\cite{Akande2010}.

\subsubsection{Response of the exchange-correlation potential}

In general, in conventional DFT the local and semi-local functionals fail in 
predicting accurate linear polarizabilities, a feature associated with the 
incorrect response of the exchange-correlation potential to the electric field. 
This failure ultimately boils down to the self-interaction
error~\cite{Korzdorfer2008,Pemmaraju2008}. Thus, it is important to 
investigate the response of our ML KS potential to the external field. The exercise 
can be simply done by plotting the potential response, namely the difference between 
the potential computed in the presence of a field and without it, 
$\Delta v = v^{\cal E}-v^{{\cal E}=0}$, as a function of the site index. At 
this point, it is useful to separate the classical and non-classical contributions 
to the exchange-correlation potential, namely to subtract from the $v_\mathrm{xc}$
previously defined the Hartree component, which is here defined as 
$(v_\mathrm{H})_i=U n_i /2$. To keep the notation simple we will keep calling
$v_\mathrm{xc}-v_\mathrm{H}$ as $v_\mathrm{xc}$.

\begin{figure}[H]
    \centering
    \includegraphics[scale = 0.61]{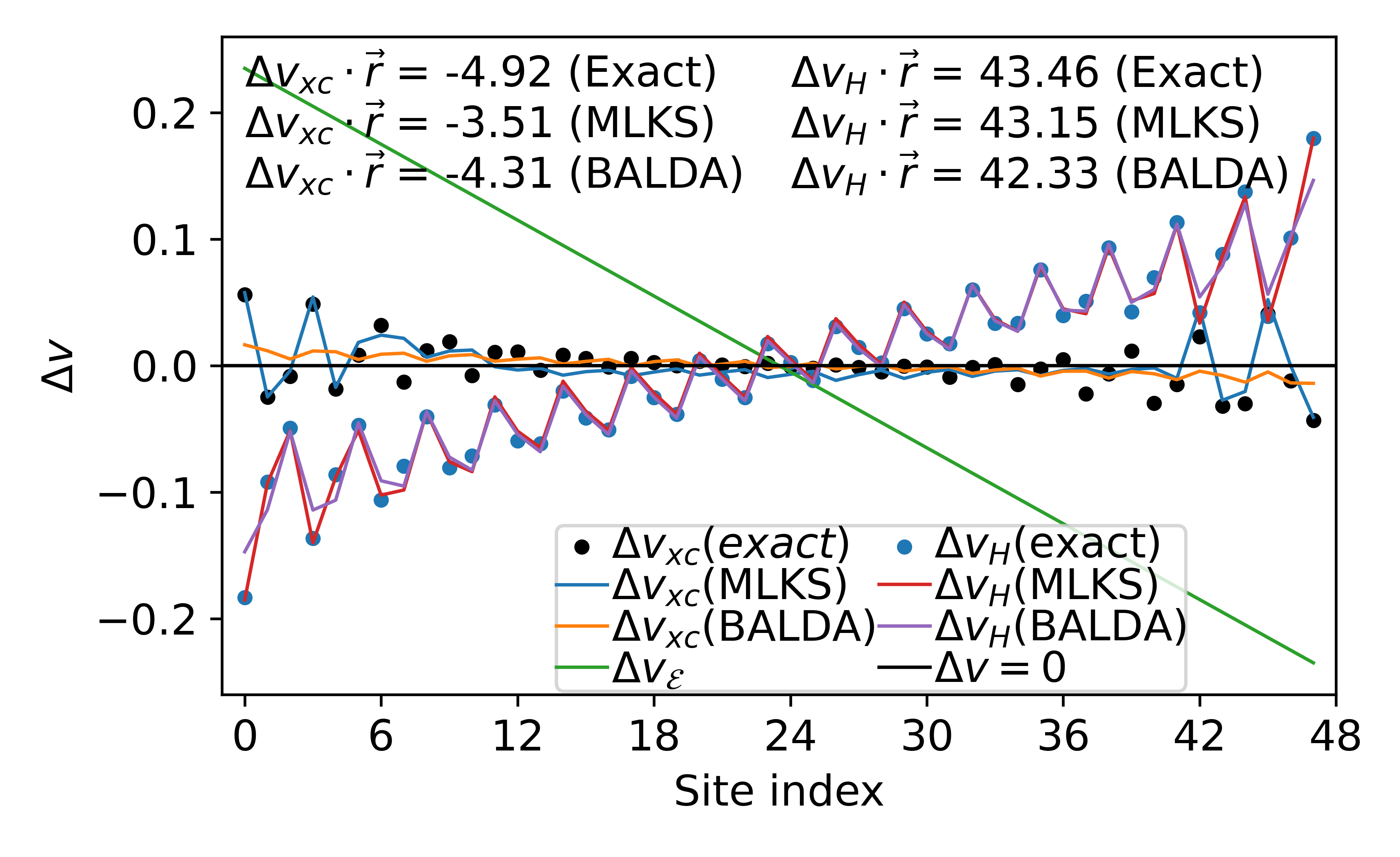}
    \caption{Response of the Hartree and exchange-correlation potential to the external
    electrostatic potential, plotted against the site index for a homogeneous $L=48$
    lattice. For the MLKS calculations we use the $a=3$ functional. A measure of the
    response can be obtained from $\Delta v \cdot \vec{r}$
    (see text), whose values are reported in the legends.}
    \label{response}
\end{figure}

In Fig.~\ref{response} we then plot $\Delta v_\mathrm{H}$, $\Delta v_\mathrm{xc}$ 
and $\Delta v_{\cal E}$, where $v_{\cal E}$ is the potential associated to the 
electric field [see Eq.~(\ref{H-Ele})]. Results are presented for a system
of $L=48$ sites and for the ML functional with $a=3$, which has returned the most 
accurate polarizability. For completeness we also include results obtained within
the BALDA. As usual, we compare the DMRG results (dots) against those obtained with 
MLKS, showing once again an excellent agreement. As expected the Hartree potential
responds against the external one. In contrast, $\Delta v_\mathrm{xc}$ has the same 
slope of $\Delta v_{\cal E}$. This behaviour is opposite to what found for 
{\it ab initio} DFT, but it is consistent with previous Bethe-ansatz LDFT
results~\cite{Akande2010}, as confirmed numerically here as well. In order to quantify 
the potential response, we compute for each of the potentials the product, 
$\sum_i \Delta v_i\cdot (i-\bar{x})$, which we refer to as $\Delta v \cdot \vec{r}$ 
and describes the slope of the response. With this definition at hand, we conclude
that the Hartree response is about ten-fold larger than the exchange-correlation ones.
It is also worth to note that the results obtained with our ML functional are
close to those obtain with the BALDA, indicating that at least for this quantity
the non-locality does not seem to play a key role.

\subsubsection{Polarizability for disordered systems as they approach the 
thermodynamics limit}
One of the major advantages of a MLKS scheme is that it can be applied to any 
system size and distribution of on-site energies (within the range of the training 
data). As such, it is ideal to investigate the effect of disorder as the
system approaches the thermodynamic limit, an exercise performed here for the
polarizability. Our analysis starts with systems comprising 21 and 42 sites,
which are accessible by DMRG. For both system sizes, we generate $n=2000$ random 
disorder configurations from $v_i^{(k)} \in [-\lambda^{(k)},\lambda^{(k)}]$ where 
$\lambda^{(k)} = 3k/2(n-1)$. Then, we compute and plot the polarizability against 
a one-dimensional measure of the disorder, namely the absolute variance 
$\delta v = \frac{1}{L}\sum_i^L |v_i|$ (note that each disorder configuration has 
zero mean). 

\begin{figure}[t]
    \centering
    \includegraphics[scale = 0.69]{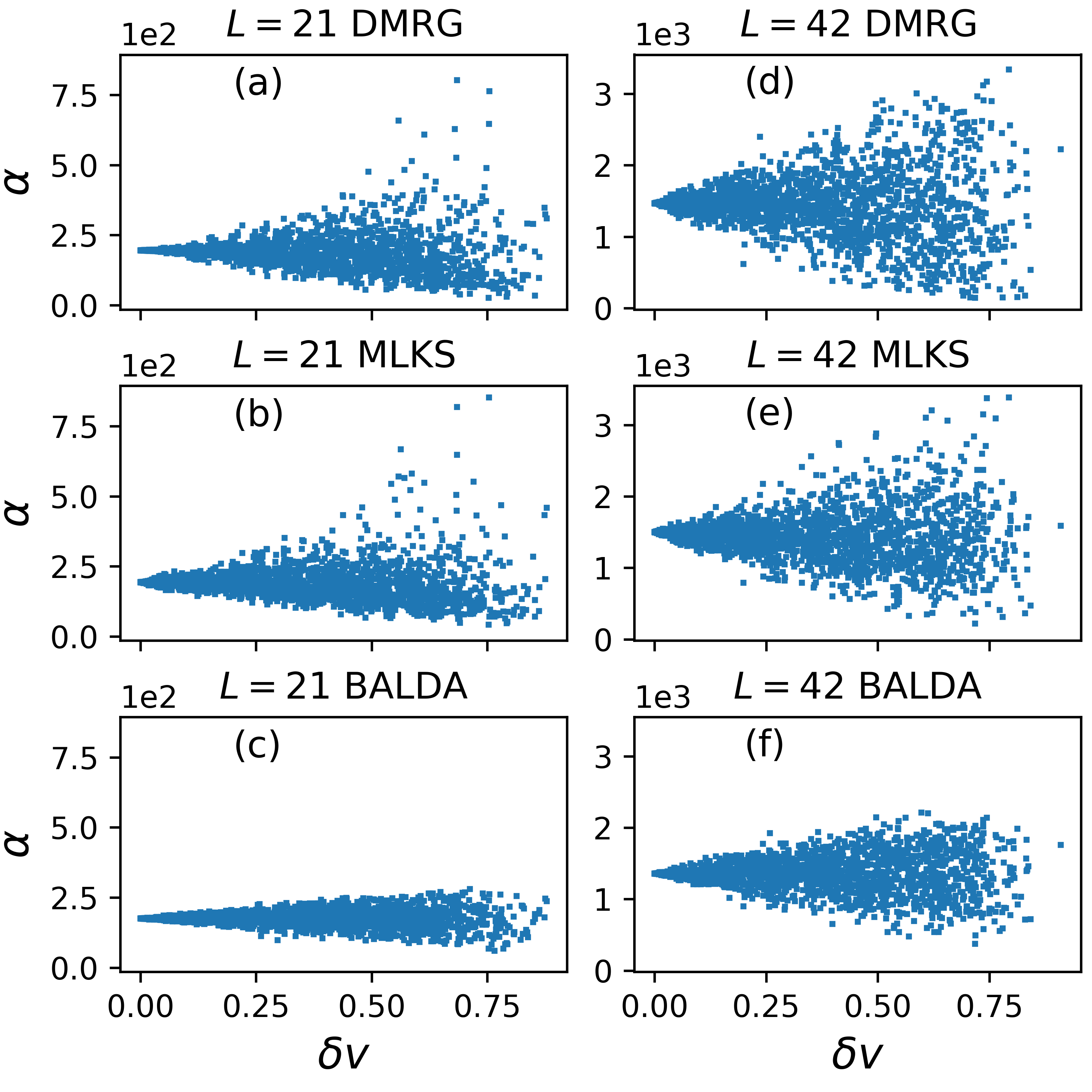}
    \caption{The polarizability, $\alpha$, against disorder strength, 
    $\delta v = \frac{1}{L}\sum_i^L |v_i|$, computed with DMRG (top row), 
    MLKS (middle row) and BALDA (bottom row), for systems of $L=21$ lattice 
    sites (left column) and $L=42$ (right column). In each sub-plot we show data
    for 2,000 disorder configurations. In the case of MLKS, we use the model 
    trained with non-locality $a=4$.}
    \label{fig:alpha_vs_disorder}
\end{figure}

The results are shown in the Fig.~\ref{fig:alpha_vs_disorder}, where we present 
the polarizability for all the 2,000 configurations, computed with DMRG
and with the KS schemes, either using our ML functional ($a=4$) or the BALDA.
Clearly, our MLKS-computed polarizabilities are in close agreement with the 
exact DMRG results across the entire range of disorder investigated. In contrast,
the BALDA provides good polarizabilities for weak disorder but tends to 
drastically under-estimate the spread of the distribution for high disorder. 
This is the situation where the site occupations may distribute over a larger 
range. It is also clear that the distributions of $\alpha$ as a function of
$\delta v$ are different for systems of different size, namely they are less
diffused for long chains. We will return on this point when analysing the limit
of large chains.

\begin{figure}[t]
    \centering
    \includegraphics[]{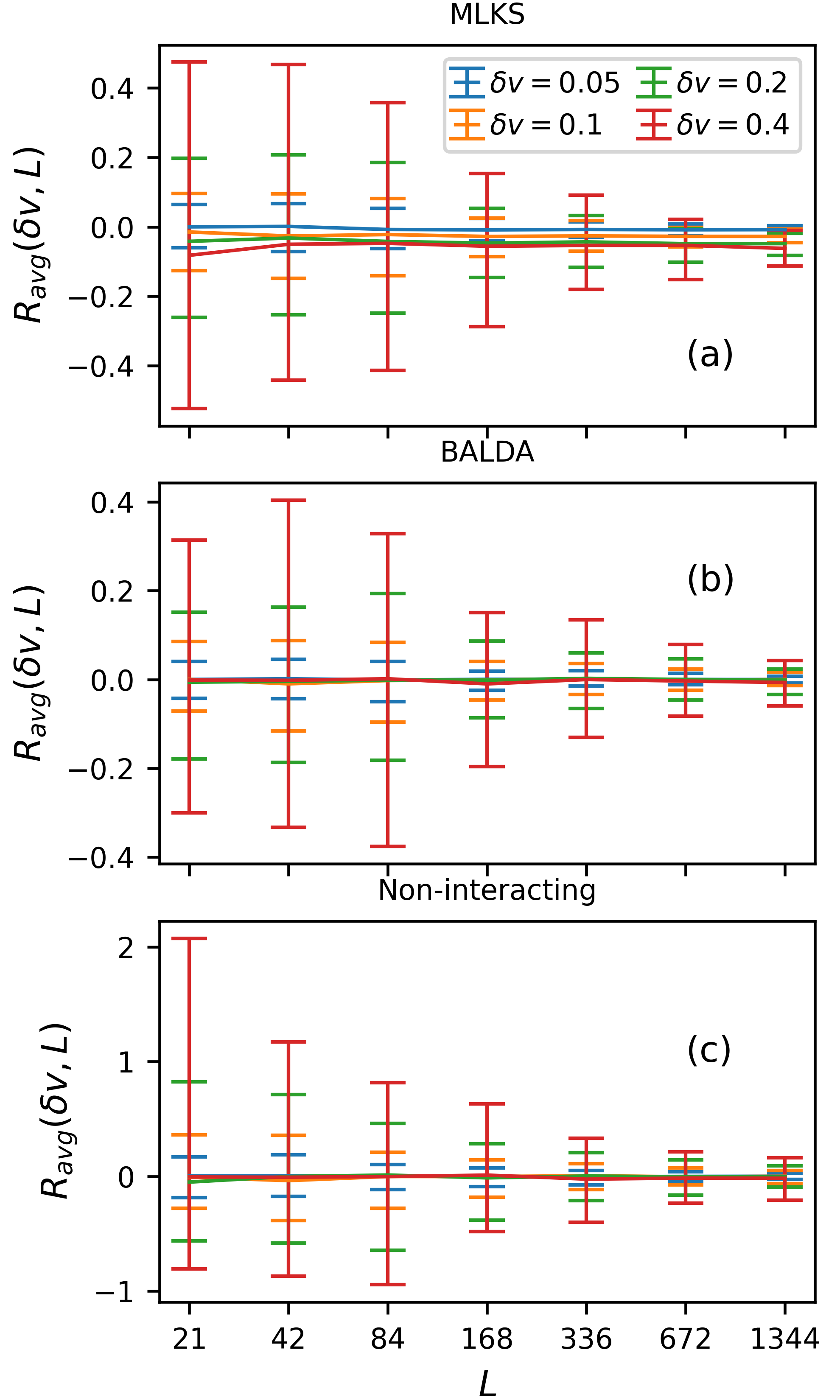}
    \caption{Relative change in the averaged polarizability, 
    $R_{\text{avg}}(\delta v, L)$ as a function of the lattice size $L$, for 
    different levels of disorder $\delta v$. The lower (upper) error bars give the 2.5\textsuperscript{th} (97.5\textsuperscript{th}) 
    percentile of the distribution. Panels (a), (b) and (c) show the results from MLKS, BALDA and 
    the non-interacting tight-binding model, respectively. There are 200 data 
    points for each $(L, \delta v)$ pair. For the MLKS scheme, we use the model 
    trained with non-locality $a=4$.}
    \label{fig:alpha_vs_L_disordered}
\end{figure}

For our large-scale investigation, we examine systems containing respectively
$L=21, 42, 84, 168, 336, 672$ and 1344 sites. In this case we consider only 
four different levels of disorder, namely $\delta v = 0.05, 0.1, 0.2, 0.4$. 
For each $L$ and $\delta v$ we generate 200 random disorder configurations 
and rescale the on-site energies so to return the required $\delta v$ value.
The polarizability is then computed according to Eq.~(\ref{alpha}) with a 
finite difference of $\mathcal{E} = 1/2L$. For this scaling exercise we monitor
the relative change in the average polarizability with respect to that of the 
homogeneous system, $\alpha_0$, namely
\begin{equation}
    \label{R-avg}
    R_{\text{avg}}(\delta v, L) = \frac{\alpha_\text{avg}(\delta v, L) - \alpha_0}{\alpha_0},
\end{equation}
while the distribution of the data is analysed by computing the 
2.5\textsuperscript{th} and 97.5\textsuperscript{th} percentiles.
%

For these system sizes we no longer have access to DMRG results, so that we limit 
the comparison of our MLKS results to those obtained with BALDA and to the 
non-interacting case. In Fig.~\ref{fig:alpha_vs_L_disordered} we present 
$R_{\text{avg}}(\delta v, L)$ against length for different levels of disorder 
and the different models used. A few key points must be noticed. Firstly, in
all cases the average polarizability remains close to that computed for the 
homogeneous case at the same level of theory. In the non-interacting case, at 
low disorder, the individual polarizabilities computed for different configurations 
distribute symmetrically about the average, a symmetry that breaks down for large 
$\delta v$. Overall then, the spread of $R_{\text{avg}}(\delta v, L)$ values
reduces monotonically with the system size, suggesting that in the thermodynamic
limit (infinite $L$) one should expect the polarizability to be that of the 
homogeneous case. 

When Coulomb repulsion is switched on (finite $U$) both our MLKS and the
BALDA scheme return us a polarizability spread significantly smaller than
that of the non-interacting case. This is not surprising since the Coulomb
repulsion is more pronounced in regions of high density, so that for the same
on-site energy random arrangement one expects the interacting problem to yield
a more homogeneous charge density. In turn, this reduces the deviation of the
computed polarizability from their average. This effect can be quantified by
looking at the difference between the 97.5\textsuperscript{th} and the 
2.5\textsuperscript{th} percentile of the distribution for the largest chain,
$L=1,344$, and strongest disorder, $\delta v = 0.4$. This is 0.1 for both MLKS
and BALDA and 0.37 for the non-interacting case. 

\begin{figure}[t]
    \centering
    \includegraphics[scale=0.097]{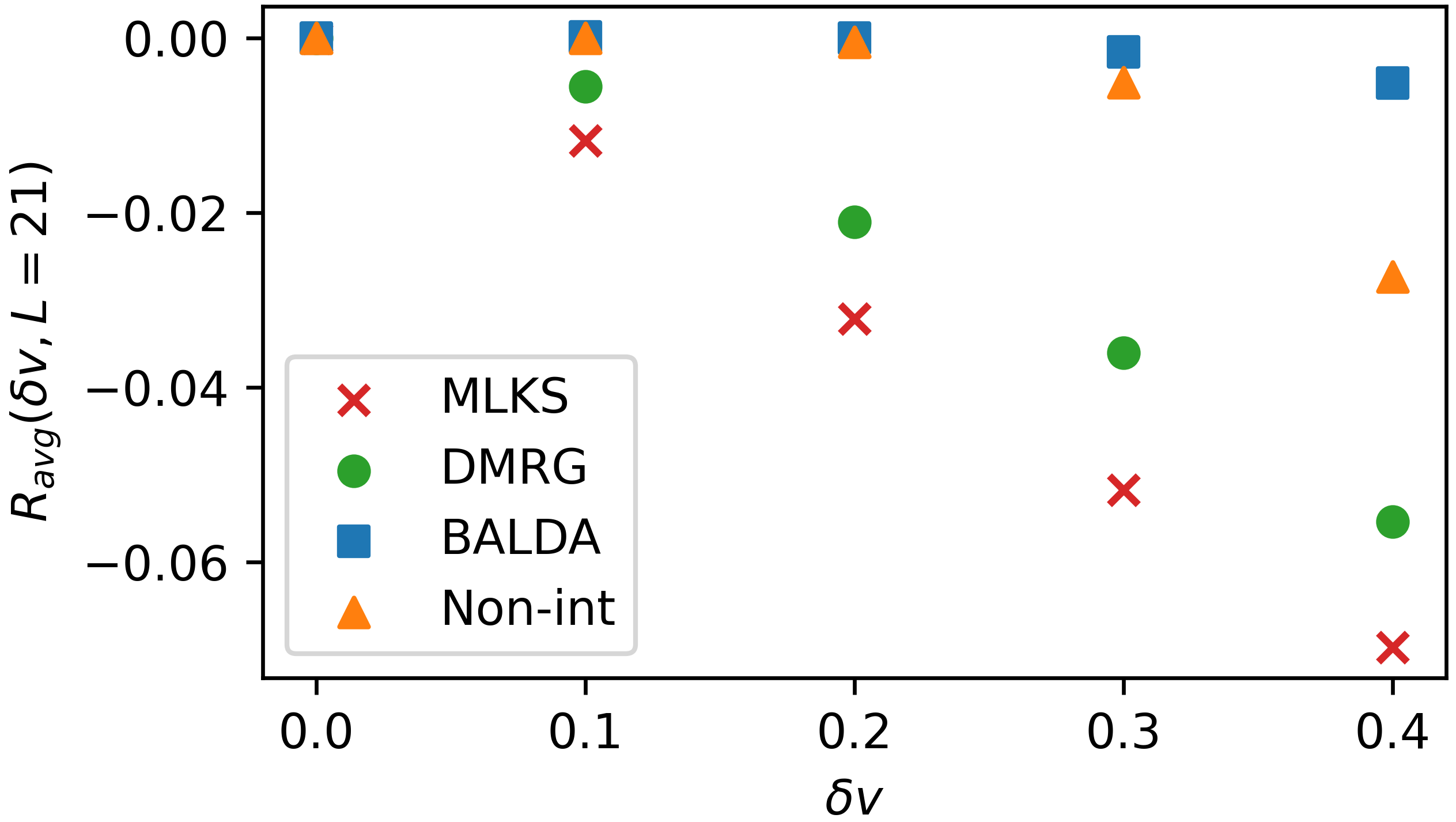}
    \caption{$R_{\text{avg}}$ against disorder strength, $\delta v = 0.1, 0.2, 0.3, 0.4$,
    for a chain containing 21 sites, computed at different levels of theory. For each 
    $\delta v$ we average over $n=16,00$ disorder configurations for MLKS and DMRG, 
    and over $n=51,200$ for BALDA and the non-interacting case. This returns 
    approximately the same variance.}
    \label{fig:r_avg}
\end{figure}

Finally, we wish to take a closer look at the dependence of the average polarizability
as a function of the disorder strength. As noted before, this eventually approaches 
the homogeneous-chain limit, but for finite chains a small dependence is noticed. 
As this is difficult to extract from the plots in Fig.~\ref{fig:alpha_vs_L_disordered}
due to the scale, in Fig.~\ref{fig:r_avg} we present the deviation of the average 
$\alpha$ from the homogeneous case as a function of $\delta v$ for a 21-site-long
chain and the different models. In constructing the graph we have averaged 
over a different number of samples, so that the variance of the different methods
is approximately similar (the average is equally statistically significant). 
Clearly the deviation from the homogeneous limit, as expected, increases with
the disorder strength. Such an increase is underestimated by both BALDA and by
the non-interacting tight-bonding model, but it is well capture by our MLKS scheme,
which remains close to the DMRG results at all $\delta v$.

\section{Conclusion and Outlook \label{conclusions}}
We have presented a method for the construction of a ML energy functional for the
one-dimensional Hubbard model, which can define a Kohn-Sham potential by 
functional derivative. This enables us to use Kohn-Sham DFT for a lattice 
model, with an accuracy comparable to that of DMRG, which is used to generate
the training set. In particular, we construct a class of functionals with
different degree of non-locality. These can be trained on systems with moderate
size and deployed for much larger ones. In general, we find that the accuracy of
the functional improves with the degree of non-locality and that most of the
error concentrates on lattice sites at the edge of the linear molecules. These
present charge distribution unevenly represented in the training set, so that the
error has to be considered numerical. All constructed functionals perform significantly
better than the Bethe-Ansatz local density approximation (BALDA), often used for the
DFT of the Hubbard model. 

Since the computational costs of our MLKS functionals are similar to that of the BALDA
and scale with the number of sites, the method can be used to investigate large
systems and/or situations where ensemble averages are important. For this reason, 
we have looked at the electrical polarizability of both homogeneous and disordered
linear chain as a function of the chain length and eventually the disorder strength. 
We find that MLKS can return polarizability values in close agreement to
DMRG results and, in the case of disorder, a very similar spread in their
distributions. This is not the case of the BALDA, which tends to under-estimate
the spreads. In any case, we find that the polarizability of disordered systems 
converges to that of the homogeneous case as the system size is increased and that
the convergence is faster for the interacting case. 

The method presented here still presents some limitations. Firstly, the functionals
are defined for a specific value of $U/t$, so that different Coulomb strengths 
will require a new parametrization. In principle, one could train a single model
to work with a range of model parameters $\{U_i,t_{ij}\}$, which effectively 
would act as a `collection' of functionals. Alternatively, one can move to 
density matrix functional theory, where the fundamental quantity becomes the
single-particle density matrix, but one has access to a more general definition
of kinetic energy~\cite{LopezSandoval2000}.
A second limitation is that here the models have been trained specifically for the 
2/3 filling case. Although we have shown that the functionals are transferable to
different, although close, filling factors, it is reasonable to expect that our
semi-local approximation will break down as the model correlation length diverges.
The Mott-insulator phase transition of the one-dimensional Hubbard model at
half filling is one such case. Despite the limitations, it appears that our MLKS
scheme can still be useful to explore similar models of relevance either at
fundamental level or for possible applications in materials science. For instance,
the extension to higher dimensions is in reach and only limited by our ability to 
generate exact results for the training. Also intriguing is the possibility to
extend our construction to the canonical ensemble~\cite{Nelson2021} and to
bosonic degrees of freedom~\cite{Bostrom2019}. One can then investigate phenomena
such as superconductivity or polaron formation and transport.

\begin{acknowledgements}
    We would like to thank Akinlolu Akande for insightful discussions regarding 
    the BALDA and DMRG. This work has been funded by the Irish Research Council 
    through a PhD scholarship (E.C.) (Grant No. GOIPG/2021/715). We acknowledge 
    Trinity Centre for High Performance Computing (TCHPC) for the provision of 
    computational resources. We acknowledge the use of GPU's provided by the 
    Nvidia Academic Hardware grant.
\end{acknowledgements}

\appendix
\section{Machine-learning the local functional \label{a=0_parity_plots_section}}

\begin{figure}[H]
    \centering
    \includegraphics[scale = 0.66]{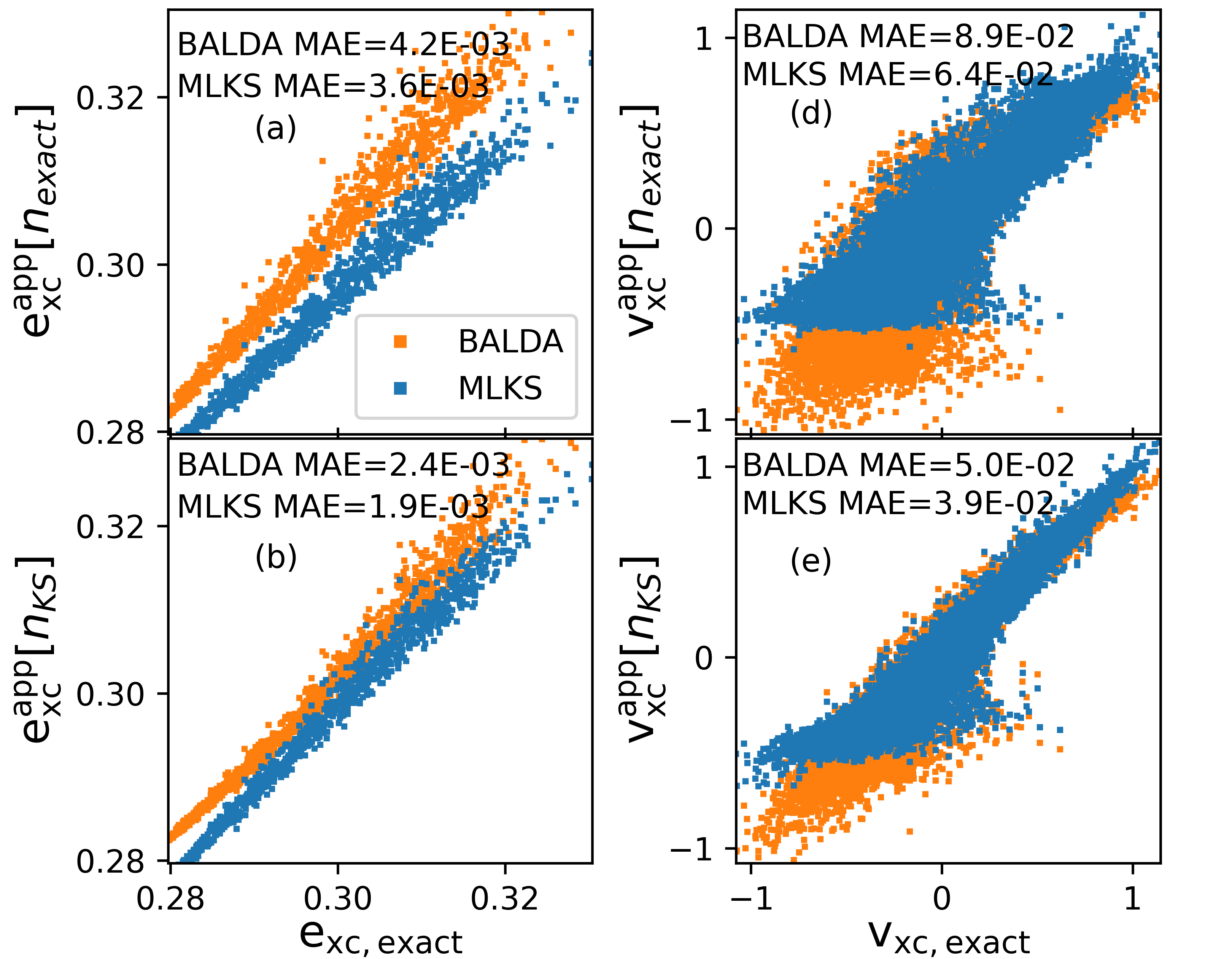}
    \includegraphics[scale = 0.625]{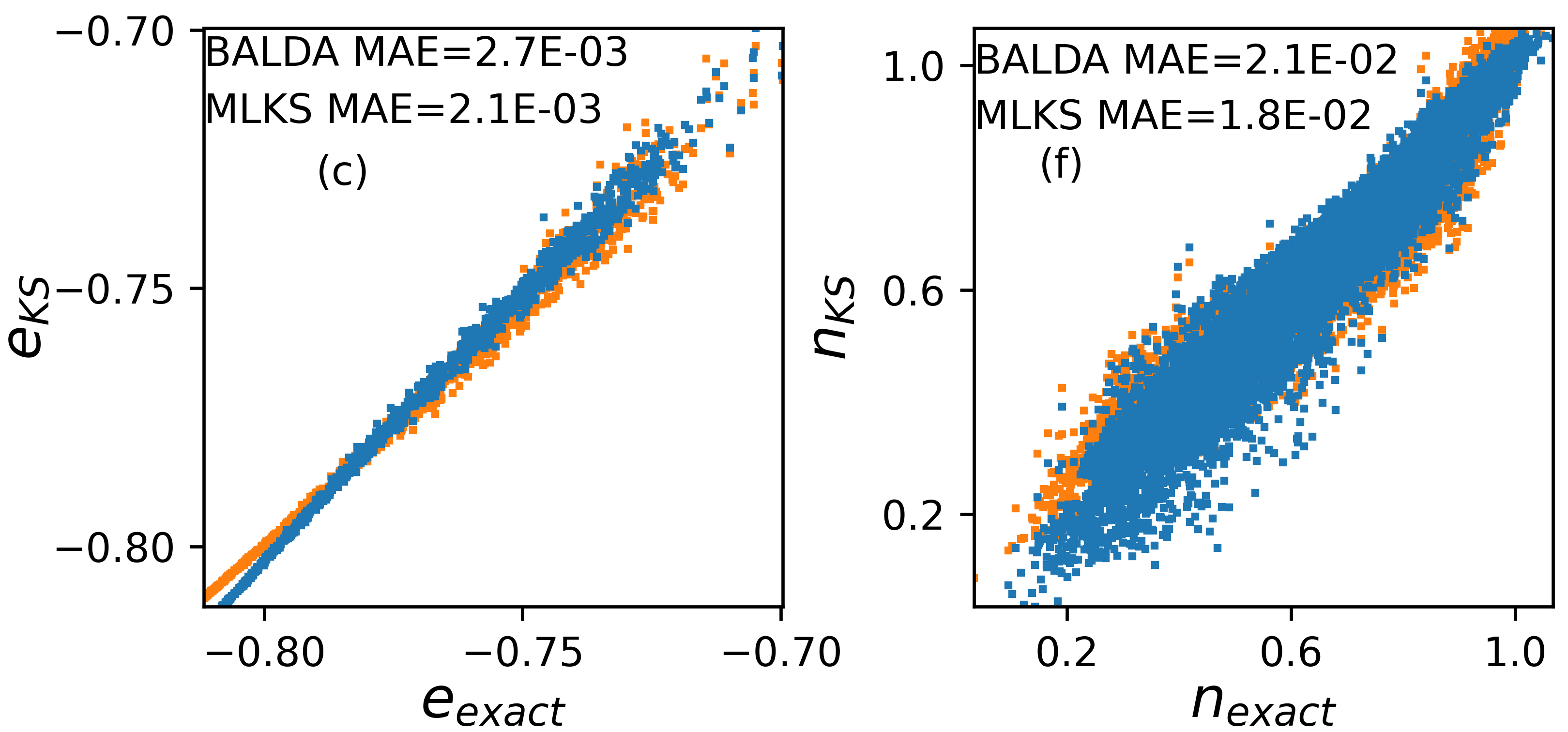}
    \includegraphics[scale = 0.625]{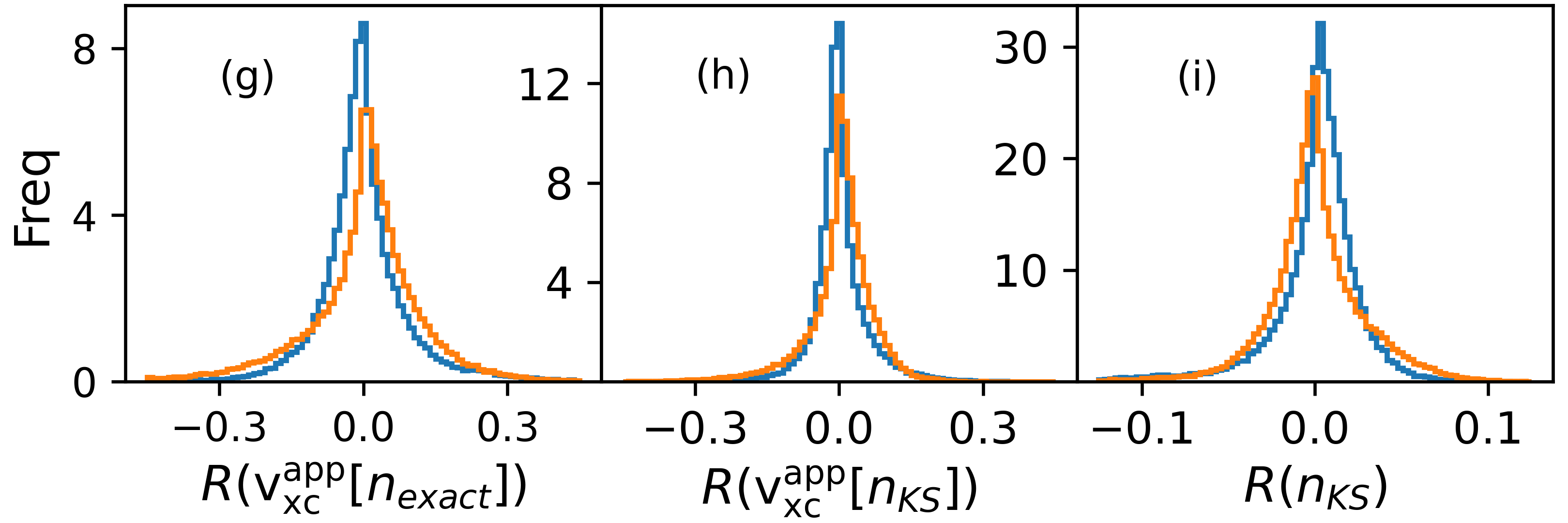}
    \caption{Analysis of the accuracy of the KS ML functional. For a system $L=60$ 
    we present the parity plot of both the ML exchange-correlation energy density, 
    $e^\mathrm{ML}_\mathrm{xc}$, and the KS exchange-correlation potential,
    $v^\mathrm{ML}_\mathrm{xc}$, computed either at the exact (DMRG) occupations [panels 
    (a) and (d), respectively] or at the converged KS occupations [panels (b) and 
    (e), respectively]. In panels (c) and (f) we present the parity plot for the
    ML KS total energy density and occupations, respectively. Results are presented 
    for the functional with a non-locality parameter $a=0$ and over a test set of 
    1,000 configurations.
    }
    \label{a0_parity_plots}
\end{figure}

In Section \ref{3a} we have examined the parity plots of both the $a=4$ functional 
as well as the BALDA. It is interesting to see how well the ML fully-local, $a=0$, 
functional compares to the BALDA. Thus, we perform the same test as in Section \ref{3a},
however with $a=0$ instead of $a=4$. The results are given in 
Fig.~\ref{a0_parity_plots}. It is interesting to see that when the ML model is 
reduced from a semi-local functional to a fully local one, it acquires some of 
the same systematic errors that are displayed by the BALDA. The $a=0$ MLKS returns 
lower MAE's for each of the six quantities [panels (a) through (f)]. This can be understood 
as follows. The BALDA is a functional that is exact in the case of the infinite
homogeneous 1D chain for any given value of the filling, whereas the $a=0$ is a 
local model that relates an input density to an energy, that is optimized to 
be most accurate for the systems in the training set (2/3 filling). In fact, any discrepancy 
between the BALDA and the MLKS $a=0$ functional implies that the latter does not recover 
the correct infinite limit of the homogeneous electron density. 

\end{document}